\documentclass[letterpaper,11pt,USenglish]{article}

\usepackage[utf8]{inputenc}
\usepackage[T1]{fontenc}
\usepackage{amsmath,amsthm,amssymb,mathrsfs}
\usepackage[margin=1in]{geometry}
\usepackage{graphicx,xcolor}
\usepackage{array,colortbl,multicol,multirow,tabularx}
\usepackage{hyperref,url}
\usepackage{enumerate,enumitem}
\usepackage[nodisplayskipstretch]{setspace} 
\usepackage{ifthen}
\usepackage{xspace}
\usepackage{calc,tikz}
\usetikzlibrary{calc}
\usepackage{multidef}
\usepackage{float}
\usepackage{xstring}
\usepackage[mathlines]{lineno}
\nolinenumbers

\newcommand{\xlen}{0.45}
\newcommand{\ylen}{0.75}
\newcommand{\hlen}{0.17}
\newcounter{ncounter}

\colorlet{vert}{green!75!black}
\colorlet{bleu}{blue!60}

\newcommand{\fatnode}[6]{
 \ifthenelse{\equal{#6}{red}}
 {\makenode{#1}{#2}{#3}{#4}{#5}{red}{red!5}{0.2}{}}
 {\makenode{#1}{#2}{#3}{#4}{#5}{black}{white}{0.2}{}}
}
\newcommand{\makenode}[9]{
 \begin{scope}[shift={(n#1)}]
 \draw[thick,draw=#6,fill=#7] (#8,-0.15) arc (-90:90:0.15) --++ (-2*#8,0) arc (90:270:0.15) -- cycle;
 \node[anchor=west] at (#86,0) {\color{#6}\scriptsize #3};
 \node[anchor=east] at (-#86,0) {\color{#6}\scriptsize #4};
 \node[anchor=south] at (0,#89) {\color{#6}\scriptsize #9};
 \node at (n#1) {\color{#6}\tiny #5}; 
 
 \ifthenelse{\equal{#2}{\sfC} \or \equal{#2}{\sfJ} \or \equal{#2}{\sfS}}
 {\node at (0,-0.405) {\scriptsize #2};}
 {\ifthenelse{\equal{#2}{\sfF}}
  {\node at (0,-0.43) {\scriptsize #2};}
  {\node[anchor=north] at (0,-0.09) {\color{#6}\scriptsize #2};}}
 \end{scope}
}

\newcommand{\ncnode}[3]{
 \draw[fill=red!10,line width=3pt,draw=red] (n#1) circle (2.1*\hlen);
 \node at (n#1) {\small$#3$};
 \node[anchor=north] at ($(n#1) + (0,0.03-2.1*\hlen)$) {\small#2};
}

\newcommand{\nenode}[3]{
 \draw[fill=red!10,line width=3pt,draw=red] (n#1) ++ (2*\hlen,\hlen) --++ (-\hlen,\hlen) --++ (-2*\hlen,0) --++ (-\hlen,-\hlen) --++ (0,-2*\hlen) --++ (\hlen,-\hlen) --++ (2*\hlen,0) --++ (\hlen,\hlen) -- cycle;
 \node at (n#1) {\small$#3$};
 \node[anchor=north] at ($(n#1) + (0,0.03-2.1*\hlen)$) {\small#2};
}

\newcommand{\arrow}[1]{
 \draw[line width=3pt,->,>=stealth] (#1*\xlen,-\ylen) --++ (2*\xlen,0);
}

\newcommand{\colorzone}[3]{
 \fill[fill=#3!15] (n#1) ++ (-0.1,0.5*\ylen) arc (90:270:0.5*\ylen) -- ($(n#2) + (0.1,-0.5*\ylen)$) arc (-90:90:0.5*\ylen) -- cycle;
}

\newcommand{\shape}[1]{
 \foreach \x in {#1}{
  \draw[thick] (n\x1)  to [bend left=10] (n\x) to [bend left=10] (n\x2);
 }
 \ifthenelse{\equal{#1}{}}{\draw[thick] (n1) to [bend left=10] (n) to [bend left=10] (n2);}{}
}
\newcommand{\buggy}[3]{
 \draw[red,line width=4pt] (n#1#2) to [bend left=#3] (n#1);
}

\newcommand\place[2]{
 \foreach \x/\c in {#2}{
  \StrLen{\c}[\y]
  \coordinate (n\c) at (#1*\xlen+\x*\xlen,-\y*\ylen);
  \IfEndWith{\c}{2}{
   \StrGobbleRight{\c}{1}[\d]
   \draw[thick] (n\d1) to [bend left=10] (n\d)  to [bend left=10] (n\d2);
  }{}
}}
\newcommand\labels[2]{
 \setcounter{ncounter}{0}
 \foreach \c/\h in {#2}{
  \csname#1node\endcsname{\c}{\h}{}{}{}
}}

\usepackage{etoolbox} 

\newcommand*\linenomathpatch[1]{%
  \cspreto{#1}{\linenomath}%
  \cspreto{#1*}{\linenomath}%
  \csappto{end#1}{\endlinenomath}%
  \csappto{end#1*}{\endlinenomath}%
}
\newcommand*\linenomathpatchAMS[1]{%
  \cspreto{#1}{\linenomathAMS}%
  \cspreto{#1*}{\linenomathAMS}%
  \csappto{end#1}{\endlinenomath}%
  \csappto{end#1*}{\endlinenomath}%
}

\expandafter\ifx\linenomath\linenomathWithnumbers
  \let\linenomathAMS\linenomathWithnumbers
  \patchcmd\linenomathAMS{\advance\postdisplaypenalty\linenopenalty}{}{}{}
\else
  \let\linenomathAMS\linenomathNonumbers
\fi

\linenomathpatch{equation}
\linenomathpatchAMS{gather}
\linenomathpatchAMS{multline}
\linenomathpatchAMS{align}
\linenomathpatchAMS{alignat}
\linenomathpatchAMS{flalign}

\theoremstyle{plain}

\theoremstyle{definition}

\newcolumntype{L}[1]{>{\raggedright\let\newline\\\arraybackslash\hspace{0pt}}m{#1}}
\newcolumntype{C}[1]{>{\centering\let\newline\\\arraybackslash\hspace{0pt}}m{#1}}
\newcolumntype{R}[1]{>{\raggedleft\let\newline\\\arraybackslash\hspace{0pt}}m{#1}}

\multidef[prefix=cal]{\ensuremath{\mathcal{#1}}\xspace}{A-Z}
\multidef[prefix=bf]{\ensuremath{\mathbf{#1}}\xspace}{A-Z}
\multidef[prefix=bf]{\ensuremath{\mathbf{#1}}\xspace}{a-z}
\multidef[prefix=sf]{\ensuremath{\mathsf{#1}}\xspace}{a-z}
\multidef[prefix=sf]{\ensuremath{\mathsf{#1}}\xspace}{A-Z}
\multidef[prefix=bb]{\ensuremath{\mathbb{#1}}\xspace}{A-Z}
\multidef[prefix=fr]{\ensuremath{\mathfrak{#1}}\xspace}{A-Z}
\multidef[prefix=scr]{\ensuremath{\mathscr{#1}}\xspace}{A-Z}

\newcommand{\key}{\mathsf{key}}
\newcommand{\dsafe}{\textsf{del}-safe\xspace}
\newcommand{\edsafe}{\textsf{edel}-safe\xspace}
\newcommand{\isafe}{\textsf{ins}-safe\xspace}
\newcommand{\dunsafe}{\textsf{del}-unsafe\xspace}
\newcommand{\edunsafe}{\textsf{edel}-unsafe\xspace}
\newcommand{\iunsafe}{\textsf{ins}-unsafe\xspace}

\title{Efficient top-down updates in AVL trees}

\author{Vincent Jugé}
\date{\begin{tabular}{c}IRIF, Université Paris Cité, CNRS \\ LIGM, Univ Gustave Eiffel, CNRS\end{tabular}}

\begin{document}


\maketitle

\begin{abstract}
Since AVL trees were invented in 1962, two major open questions about rebalancing operations, which found positive answers in other balanced binary search trees, were left open:
can these operations be performed top-down (with a fixed look-ahead), and can they use an amortised constant number of write operations per update?
We propose an algorithm that answers both questions positively.
\end{abstract}

%



\section{Introduction}\label{sec:intro}

Balanced search trees are among the most fundamental and basic data structures in computer science.
Their formidable story started with the invention of AVL trees by Adel'son-Vel'skii and Landis~\cite{AVL62} in 1962, who proposed a data structure and a maintenance algorithm thanks to which the insertion, deletion and research of an element in a linearly ordered set of size~$n$ was feasible in time~$\calO(\log(n))$.

AVL trees are binary search trees in which each node~$x$ maintains some form of \emph{balance}: the heights of its children differ from each other by at most~$1$.
Satisfying this property at each node ensures that the tree is of height~$\calO(\log(n))$.
The number of comparisons required to search an element in a binary search tree being linear in the height of the tree, it is thus logarithmic in~$n$.

However, for instance, inserting a node into the sub-tree rooted at the left child of our node~$x$ may increase the height of that sub-tree, damaging the balance of~$x$.
Thus, the challenge of maintenance algorithms consists in efficiently modifying the parenthood relations inside the tree in order to ensure that each node remains balanced.

Unfortunately, in spite of their excellent worst-case height, AVL trees suffer from updating algorithms that are significantly less efficient than those of other balanced binary search tree structures such as weight-balanced trees~\cite{NR72}, red-black trees~\cite{RB78}, half-balanced trees~\cite{O82}, splay trees~\cite{Splay85} or, more recently, weak AVL trees~\cite{HST15}.
Indeed, as illustrated in Table~\ref{table:stats}, they have two shortcomings, from which almost all other algorithms are exempt:
\begin{enumerate}[label=$\sfS_\arabic*$:]
\item starting form an empty tree, performing~$n$ updates in a row may trigger~$\Theta(n \log(n))$ rotations~\cite{ALT16} or, more generally, write operations, which may also consist in modifying the balance of a node without changing the tree structure;

\item updates following an insertion or deletion must be performed in a bottom-up fashion;
the inventors of weak AVL trees even said, in~\cite[p.~23]{HST15}, that ``\hspace{0.2pt}Top-down insertion or deletion with fixed look-ahead in an AVL tree is problematic.
(We do not know of an algorithm; we think there is none.)''
\end{enumerate}

{\renewcommand*{\arraystretch}{1.1}
\begin{table}[t]
\begin{center}
\begin{tabular}{|C{37mm}|L{0mm}C{25mm}R{0mm}|L{0mm}C{25mm}R{0mm}|L{0mm}C{25mm}R{0mm}|}
\hline
\multirow{2}{*}{Tree family} & \multicolumn{3}{c|}{Worst-case} & \multicolumn{3}{c|}{Worst-case amortised} & \multicolumn{3}{c|}{Top-down update}\\
 & \multicolumn{3}{c|}{height} & \multicolumn{3}{c|}{write cost/update} & \multicolumn{3}{c|}{algorithm} \\
\hline
AVL (state of the art) & & \multirow{2}{*}{$1.44 \log_2(n)$} & \multirow{2}{*}{\llap{\cite{AVL62}}} & & {}$\Theta(\log(n))$ & \llap{\cite{ALT16}} & & no & \\
\cline{1-1}
\cline{5-10}
AVL (this article) & & & & & {}$\Theta(1)$ & & & yes & \\
\hline
Weight-balanced & & {}$2 \log_2(n)$ & \llap{\cite{NR72}} & & {}$\Theta(1)$ & \llap{\cite{BM80}} & & yes & \llap{\cite{VJ24b,LW93}} \\
\hline
Red-black & & {}$2 \log_2(n)$ & \llap{\cite{RB78}} & & {}$\Theta(1)$ & \llap{\cite{RB78}} & & yes & \llap{\cite{RB78}} \\
\hline
Half-balanced & & {}$2 \log_2(n)$ & \llap{\cite{O82}} & & {}$\Theta(1)$ & \llap{\cite{O82}} & & for insertions & \llap{\cite{O80}} \\
\hline
Splay & & {}$n$ & \llap{\cite{Splay85}} & & {}$\Theta(1)$ & \llap{\cite{Splay85}} & & yes & \llap{\cite{Splay85}} \\
\hline
Weak AVL & & {}$2 \log_2(n)$& \llap{\cite{HST15}} & & {}$\Theta(1)$ & \llap{\cite{HST15}} & & yes & \llap{\cite{HST15}}\\
\hline
\end{tabular}
\end{center}
\caption{Asymptotic approximate worst-case height, amortised number of rewrite operations per update with~$n$ nodes, and existence of top-down update algorithms.
Next to each piece of information are indicated references where it is proved.}
\label{table:stats}
\end{table}
}

Requiring only~$\Theta(1)$ write operations per update instead of~$\Theta(\log(n))$, even in an amortised fashion, would clearly improve the efficiency of AVL trees.
As mentioned in~\cite{RB78,HST15}, being able to use top-down algorithms is also important, because requiring a bottom-up pass is costly:
either it forces each node to maintain a link toward its parent, or it must be performed \emph{via} recursive calls or storing visited nodes in a stack.
This second approach is even more expensive in distributed settings, because each process should lock the entire branch it is visiting, preventing any update from starting before the previous update has ended.

In this article, we prove that top-down insertion and deletion with fixed look-ahead in an AVL tree \emph{is} feasible\footnote{Up to keeping a pointer on an internal node we may wish to delete, as in top-down algorithms for other data structures.}
and can be write-efficient.
We proceed step-by-step:
in Section~\ref{sec:sota}, we briefly present textbook algorithms for updating AVL trees;
in Section~\ref{sec:BU}, we propose a bottom-up algorithm that requires~$\calO(1)$ write operations per update in an amortised fashion;
in Section~\ref{sec:td}, we propose a top-down algorithm;
finally, in Section~\ref{sec:TD}, we combine the advantages of the two previous algorithms, and propose a top-down algorithm that requires~$\calO(1)$ write operations per update in an amortised fashion.

\section{State-of-the-art}
\label{sec:sota}

\subsection{Preliminaries}
\label{sec:prelim}

A \emph{binary search tree} is a plane rooted tree~$\calT$ in which each tree node~$x$ has one \emph{key}~$\key(x)$, one \emph{left child}~$x_1$ and one \emph{right child}~$x_2$, which can be \emph{tree nodes} or \emph{null nodes};
null nodes represent the empty tree and have no key nor children, and tree nodes whose two children are null nodes are called \emph{leaves}.
By definition, the \emph{descendants} of~$x$ consist of~$x$, its children~$x_1$ and~$x_2$, and all their descendants (if they are tree nodes); these descendants form a sub-tree of~$\calT$ rooted at~$x$ and denoted by~$\calT(x)$, and~$x$ is one of their \emph{ancestors}.
Below, we may often informally identify each node~$x$ with the tree~$\calT(x)$, e.g., say that~$\calT(x_1)$ is a child of~$x$.

Node keys are pairwise distinct, and must belong to a linearly ordered set.
For each tree node~$y$ that descends from~$x$, we have~$\key(y) < \key(x)$ if~$y$ belongs to~$\calT(x_1)$, and~$\key(y) > \key(x)$ if~$y$ belongs to~$\calT(x_2)$.
Moreover, the \emph{height} of~$x$ is the length (i.e., the number of parent-child edges) of the longest downward path starting from~$x$ and ending in a leaf; it is denoted by~$\sfh(x)$.
By convention, each null node has height~$-1$.

A binary search tree is an AVL tree when the height difference~$\sfh(x_1) - \sfh(x_2)$ belongs to the set~$\{-1,0,1\}$ for all tree nodes~$x$.
This ensures that an AVL tree with~$n$ nodes is of height~$\calO(\log(n))$.

In practice, AVL tree structures are implemented by assigning a \emph{rank}~$\sfr(x)$ to each tree node~$x$, and storing differences between the ranks of~$x$ and its children~$x_1$ and~$x_2$; the rank of a tree is defined as the rank of its root.
Ranks are aimed to coincide with node heights, which happens when~$\sfr(x) = \max\{\sfr(x_1),\sfr(x_2)\}+1$ for all tree nodes~$x$, and~$\sfr(y) = -1$ for all null nodes~$y$.

Hence, we should control the rank differences~$\sfr(x) - \sfr(x_i)$ for each node~$x$.
Focusing on these differences, we say that~$x_i$ is an~$\ell$-child if~$\sfr(x) - \sfr(x_i) = \ell$, and that~$x$ is a~$\{\ell,\ell'\}$-node if its children are~$\ell$- and~$\ell'$-children.

Since binary search trees represented ordered sets, the three main operations they must support are searching, inserting, and deleting an element.
Searching an element is easy: when searching a key~$k$ in a non-empty tree~$\calT$, we compare~$k$ with~$\key(t)$; then, we keep searching~$k$ in~$\calT(x_1)$ if~$k < \key(t)$, or in~$\sfR(t)$ if~$k > \key(t)$, or we declare that~$k$ was found if~$k = \key(t)$; whereas, when searching~$k$ in an empty tree, we must declare that~$k$ was nowhere to be found.

Inserting or deleting an element is substantially more difficult, because it requires altering the structure of the tree: this may change its height and even damage the balance of some tree nodes.
As a result, the relation~$\sfr(x) = \max\{\sfr(x_1),\sfr(x_2)\}+1$ might fail for one node~$x$, or one rank difference~$\sfr(x_1) - \sfr(x_2)$ might no longer belong to the set~$\{-1,0,1\}$.
The purpose of bottom-up rebalancing algorithms is precisely to eliminate this anomaly, either directly, or by ``propagating it upward'', i.e., making it concern a strict ancestor of~$x$, with rank larger than~$\sfr(x)$.

\subsection{Tree cuts and cut rebalancing}
\label{sec:tree-cut}

A \emph{cut} of a tree~$\calT$ is a list~$\scrT$ of sub-trees~$\calT^1,\calT^2,\ldots,\calT^\ell$ of~$\calT$, ordered from left to right, that is maximal for inclusion;
when~$x$ and~$y$ belong to sub-trees~$\calT^i$ and~$\calT^j$ such that~$i < j$, we must have~$\key(x) < \key(y)$.
By construction, exactly~$\ell-1$ nodes of~$\calT$, denoted by~$a^1,a^2,\ldots,a^{\ell-1}$, are strict ancestors of at least one sub-tree~$\calT^i$, and the in-order enumeration of~$\calT$ consists in enumerating nodes in~$\calT^1$, then~$a^1$, then nodes in~$\calT^2$, then~$a^2$, and so on, until we enumerate the node~$a^{\ell-1}$ and the nodes in~$\calT^\ell$;
these~$\ell-1$ nodes~$a^i$ are called the \emph{ancestors} of the cut, each node~$a^i$ being the least common ancestor of~$\calT^i$ and~$\calT^{i+1}$.
This cut may also be denoted by~$\scrT = \calT(x^1,x^2,\ldots,x^\ell)$, where each~$x^i$ is the root of the tree~$\calT^i$.

Below, we may modify the structure of a tree~$\calT$, thus generalising so-called \emph{rotations}:
starting from a cut~$\scrT = (\calT^1,\calT^2,\ldots,\calT^\ell)$ of~$\calT$ and its ancestors~$a^1,a^2,\ldots,a^{\ell-1}$, we may reset the children of each node~$a^i$, and decide that the new root should be some node~$a^j$, as long as the graph we obtain is a binary search tree rooted at~$a^j$.
While doing so, (i)~we must never modify the structure or rank of the sub-trees~$\calT^i$, and (ii)~node ranks of nodes~$a^i$ are reassigned to make sure that each node~$a^i$ is a~$\{1,1\}$- or a~$\{1,2\}$-node in the resulting tree.
Such an operation is called a \emph{cut rebalancing}, and is easy to describe diagrammatically, like in Figure~\ref{fig:balancing-0} below.

\subsection{Textbook bottom-up algorithm}
\label{sec:textbook}

Inserting a key may create a \emph{zero-edge}, i.e., an edge between a parent~$x$ and its child~$x_i$ such that~$\sfr(x) = \sfr(x_i)$;
in that case, the integer~$\sfr(x) = \sfr(x_i)$ is called the \emph{rank} of the zero-edge.
A zero-edge is created when~$x$ is a leaf below which we wish to insert the key~$k$, because some child~$x_i$ of~$x$, which used to be null, becomes a leaf.

Similarly, deleting a key may create a~$\{2,2\}$- or~$\{1,3\}$-node, which we call a \emph{four-node};
this happens if~$x$ is a~$\{1,2\}$-node one of whose children is a leaf containing the key~$k$ we wish to delete, because that leaf will be transformed into a null node.

Zero-edges are dealt with as follows, and as illustrated in Figure~\ref{fig:balancing-0}.
Up to using a left-right symmetry, and without loss of generality, we assume that~$\sfr(x_1) \geqslant \sfr(x_2)$, i.e., that bot~$\sfr(x)$ and~$\sfr(x_1)$ are equal to the rank~$\sfr$ of the zero-edge;
for each descendant~$x_u$ of~$x$, the height difference~$\sfr - \sfr(x_u)$ is denoted by~$\delta_u$:

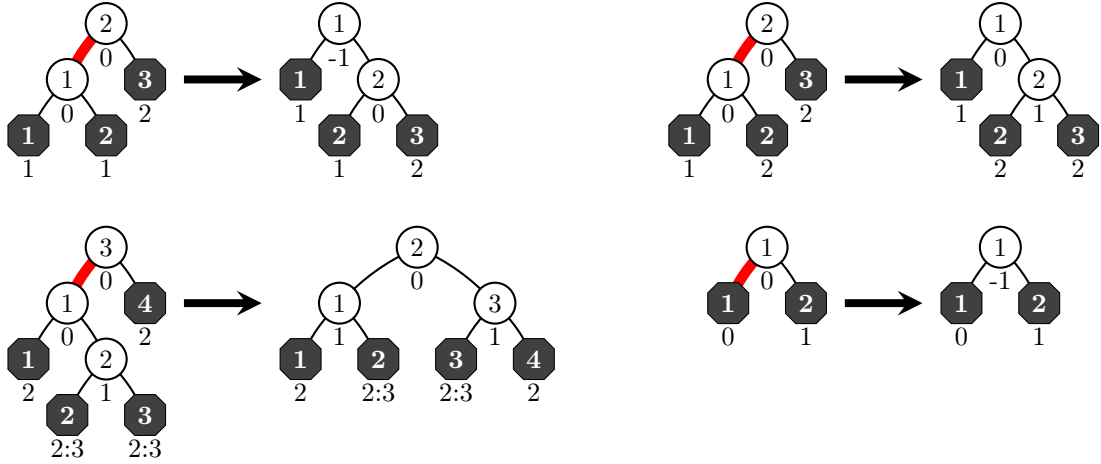
\begin{figure}[t]
\begin{center}
\begin{tikzpicture}[scale=0.8]
\renewcommand{\xlen}{0.646}
\renewcommand{\ylen}{0.93}
\place{0}{2/,1/1,0/11,2/12,3/2}
\buggy{}{1}{10}
\labels{na}{1/0,/0}
\labels{nb}{11/1,12/1,2/2}

\arrow{4}

\place{7}{1/,0/1,2/2,1/21,3/22}
\labels{na}{/-1,2/0}
\labels{nb}{1/1,21/1,22/2}


\place{17}{2/,1/1,0/11,2/12,3/2}
\buggy{}{1}{10}
\labels{na}{1/0,/0}
\labels{nb}{11/1,12/2,2/2}

\arrow{21}

\place{24}{1/,0/1,2/2,1/21,3/22}
\labels{na}{/0,2/1}
\labels{nb}{1/1,21/2,22/2}


\begin{scope}[shift={(0,-4*\ylen)}]
\place{0}{2/,1/1,0/11,2/12,1/121,3/122,3/2}
\buggy{}{1}{10}
\labels{na}{1/0,12/1,/0}
\labels{nb}{11/2,121/2:3,122/2:3,2/2}

\arrow{4}

\place{7}{3/,1/1,0/11,2/12,5/2,4/21,6/22}
\labels{na}{1/1,/0,2/1}
\labels{nb}{11/2,12/2:3,21/2:3,22/2}


\place{18}{1/,0/1,2/2}
\buggy{}{1}{10}
\labels{na}{/0}
\labels{nb}{1/0,2/1}

\arrow{21}

\place{24}{1/,0/1,2/2}
\labels{na}{/-1}
\labels{nb}{1/0,2/1}
\end{scope}
\end{tikzpicture}
\vspace{-1.2em}
\end{center}
\caption{Eliminating or propagating upward a (thick, red-painted) zero-edge of rank~$\sfr$ when~$\delta_{11} = 1$ and~$\delta_2 = 2$ (row~1), or~$\delta_{11} = \delta_2 = 2$ (row~2, left), or~$\delta_2 = 1$ (row~2, right).
Each black octagon labelled~$i$ represents the root of the sub-tree~$\calT^i$;
each white circle labelled~$i$ represents the cut ancestor~$a^i$.
After the cut rebalancing operation has taken place, the rank of a node~$a^i$ is defined as the smallest integer larger than the ranks of its children.
Below each node~$z$ is written the rank difference~$\sfr - \sfr(z)$; when two rank differences~$\delta$ and~$\delta'$ are possible, we just write~$\delta:\delta'$.
\label{fig:balancing-0}}
\end{figure}

\begin{enumerate}[leftmargin=4.8mm,itemsep=0.5pt,topsep=0.5pt]
\item If~$\delta_{11} = \delta_{12} = 1$ and~$\delta_2 = 2$ (row~1, left), we rebalance the cut~$\scrT = \calT(x_{11},x_{12},x_2)$ by letting the cut ancestors~$a^1 = x_1$ and~$a^2 = x$ be the parents of~$\calT^1$ and~$a^2$, and~$\calT^2$ and~$\calT^3$, respectively;
hence,~$a^1$ is the root of the resulting tree.
This is a \emph{simple rotation}.
\label{sota:case-1}

\item If~$\delta_{11} = 1$ and~$\delta_{12} = \delta_2 = 2$ (row~1, right), we perform the same rotation; however, the resulting ranks differ.
\label{sota:case-2}

\item If~$\delta_{12} = 1$ and~$\delta_{12} = \delta_2 = 2$ (row~2, left), we rebalance the cut~$\scrT = \calT(x_{11},x_{121},x_{122},x_2)$ to give it the shape of a complete binary tree: we let the cut ancestors~$a^1$,~$a^2$ and~$a^3$ be the parents of~$\calT^1$ and~$\calT^2$,~$a^1$ and~$a^3$, and~$\calT^3$ and~$\calT^4$, respectively;
hence,~$a^2$ is the root of the resulting tree.
This is a \emph{double rotation}.
\label{sota:case-3}

\item If~$\delta_2 = 1$ (row~2, right), we rebalance the cut~$\scrT = \calT(x_1,x_2)$ by not changing its structure, but simply increasing the rank of the cut ancestor~$a^1 = x$: this is a \emph{promotion}.
\label{sota:case-4}
\end{enumerate}

Doing so, whenever we face a zero-edge~$e$ of rank~$\sfr$, we will either eliminate it at once or replace it by a new zero-edge~$e'$ of rank~$\sfr+1$, placed just above where~$e'$ used to be.
Consequently, if inserting a key creates a zero-edge, we will simply propagate that zero-edge upward until it disappears.

Below, all transformations will be organised similarly, by describing a cut we rebalance, and referring to a particular diagram in which rank differences are explicitly written.
In particular, we will omit describing explicitly in plain text which cut ancestor becomes the parent of which nodes, this information being already visible in the diagrams we will refer to.

Likewise, four-nodes are dealt with as follows, and as illustrated in Figure~\ref{fig:balancing-1}.
Up to using a left-right symmetry, we assume that~$\sfr(x_1) \geqslant \sfr(x_2)$;
for each descendant~$x_u$ of~$x$, the height difference~$\sfr - \sfr(x_u)$ is denoted by~$\delta_u$:

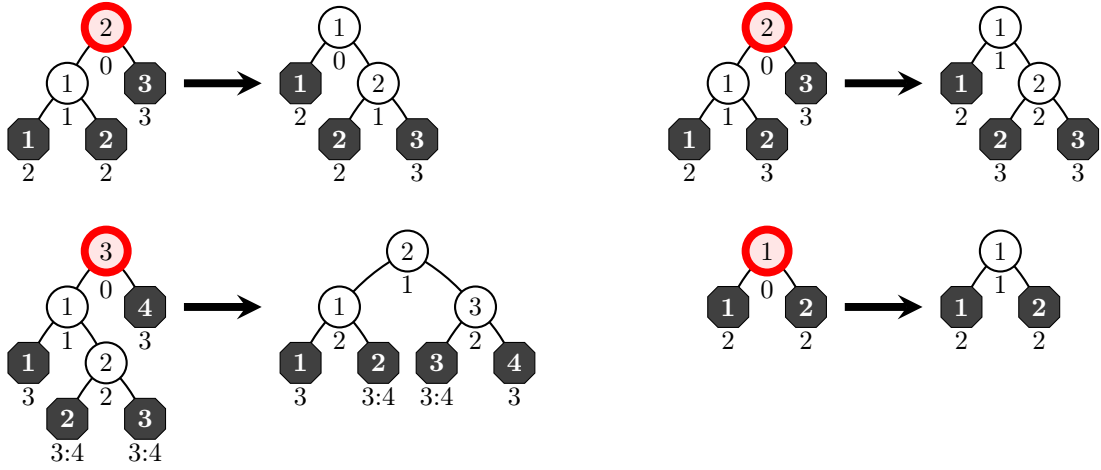
\begin{figure}[t]
\begin{center}
\begin{tikzpicture}[scale=0.8]
\renewcommand{\xlen}{0.646}
\renewcommand{\ylen}{0.93}
\place{0}{2/,1/1,0/11,2/12,3/2}
\labels{na}{1/1}
\ncnode{}{0}{2}
\labels{nb}{11/2,12/2,2/3}

\arrow{4}

\place{7}{1/,0/1,2/2,1/21,3/22}
\labels{na}{/0,2/1}
\labels{nb}{1/2,21/2,22/3}


\place{17}{2/,1/1,0/11,2/12,3/2}
\labels{na}{1/1}
\ncnode{}{0}{2}
\labels{nb}{11/2,12/3,2/3}

\arrow{21}

\place{24}{1/,0/1,2/2,1/21,3/22}
\labels{na}{/1,2/2}
\labels{nb}{1/2,21/3,22/3}


\begin{scope}[shift={(0,-4*\ylen)}]
\place{0}{2/,1/1,0/11,2/12,1/121,3/122,3/2}
\labels{na}{1/1,12/2}
\ncnode{}{0}{3}
\labels{nb}{11/3,121/3:4,122/3:4,2/3}

\arrow{4}

\place{7}{2.75/,1/1,0/11,2/12,4.5/2,3.5/21,5.5/22}
\labels{na}{1/2,/1,2/2}
\labels{nb}{11/3,12/3:4,21/3:4,22/3}


\place{18}{1/,0/1,2/2}
\ncnode{}{0}{1}
\labels{nb}{1/2,2/2}

\arrow{21}

\place{24}{1/,0/1,2/2}
\labels{na}{/1}
\labels{nb}{1/2,2/2}
\end{scope}
\end{tikzpicture}
\vspace{-1.2em}
\end{center}
\caption{Eliminating or increasing the rank of a (thick, red-painted) four-node when~$\delta_{11} = 2$ and~$\delta_2 = 3$ (row~1), or~$\delta_{11} = \delta_2 = 3$ (row~2, left), or~$\delta_1 = \delta_2 = 2$ (row~2, right).
\label{fig:balancing-1}}
\end{figure}

\begin{enumerate}[leftmargin=4.8mm,itemsep=0.5pt,topsep=0.5pt]
\setcounter{enumi}{4}

\item If~$\delta_{11} = \delta_{12} = 2$ and~$\delta_2 = 3$ (row~1, left), we rebalance the cut~$\scrT = \calT(x_{11},x_{12},x_2)$ by performing a simple rotation.
\label{sota:case-5}

\item If~$\delta_{11} = 2$ and~$\delta_{12} = \delta_2 = 3$ (row~1, right), we perform the same rotation; the resulting ranks differ.
\label{sota:case-6}

\item If~$\delta_{12} = 2$ and~$\delta_{12} = \delta_2 = 3$ (row~2, left), we rebalance the cut~$\scrT = \calT(x_{11},x_{121},x_{122},x_2)$ by performing a double rotation.
\label{sota:case-7}

\item If~$\delta_2 = 2$ (row~2, right), we rebalance the cut~$\scrT = \calT(x_1,x_2)$ by not changing its structure, but simply decreasing the rank of the cut ancestor~$a^1 = x$: this is a \emph{demotion}.
\label{sota:case-8}
\end{enumerate}

Cut rebalancing operations~\ref{sota:case-5} to~\ref{sota:case-7} are the same as operations~\ref{sota:case-1} to~\ref{sota:case-3} for eliminating zero-edges, except that the ranks of~$x_1$,~$x_2$ and their descendants were all shifted down by one.
In all eight cases, these operations result either in eliminating the four-node~$x$ or  in making the (former) parent of~$x$ our new four-node.

\section{Efficient bottom-up updates}
\label{sec:BU}

\subsection{Overview and complexity analysis}
\label{sec:overview-BU}

The textbook algorithm presented in Section~\ref{sec:textbook} aims at eliminating or propagating upward zero-edges and four-nodes.
Thus, each insertion or deletion update consists in (i)~finding the location at which we want to create or delete a tree leaf, (ii)~observe that we may have created a zero-edge or four-node, (iii)~moving this anomaly upward, until (iv)~we eliminate it once and for all.
Step~(i) requires no write operation on the tree structure, and steps (ii) and~(iv) occur once per update.
However, step (iii) may require arbitrarily many write operations per update.

Hence, cut rebalancing operations of cases~\ref{sota:case-1} to~\ref{sota:case-8} are called \emph{terminal} if they result in eliminating the anomaly (this is case~(iv)), and \emph{transient} if they just propagate it upward (this is case~(iii)).
We should prove that updating~$n$ times an initially empty AVL tree triggers~$\calO(n)$ transient operations.

With the textbook algorithm, this statement we wish to prove is invalid~\cite{ALT16}.
Consequently, we will tweak the operations performed in cases~\ref{sota:case-1} to~\ref{sota:case-8}, aiming at making them terminal whenever possible.

The complexity proof of the resulting algorithm is based on \emph{potential} techniques, i.e., on studying the variations of two quantities: the number of~\emph{$3$-full} nodes, a notion we define next, and the number of~$\{1,2\}$-nodes.
As we will prove:
\begin{enumerate}[leftmargin=5.8mm,itemsep=0.5pt,topsep=0.5pt]
\item each update or write operation modifies the neighbourhood (parenthood relations and/or rank) of a constant number of nodes, hence it increases both these quantities by at most a constant;
\label{statement-1}

\item each transient operation aimed at eliminating a zero-edge of rank~$\sfr \geqslant 4$ decreases the number of~$3$-full nodes;
\label{statement-2}

\item each transient operation triggered by a deletion decreases the number of~$\{1,2\}$-nodes, and does not increase the number of~$3$-full nodes.
\label{statement-3}
\end{enumerate}

Hence, if~$n$ updates trigger~$\sfI$ transient insertion-related and~$\sfD$ transient deletion-related operations, the numbers of~$3$-full nodes and~$\{1,2\}$-nodes in resulting tree are bounded from below by~$0$, and from above by~$\calO(n) - \sfI$ and~$\calO(n+\sfI) - \sfD$, respectively.
It follows that~$\sfI = \calO(n)$ and that~$\sfD = \calO(n+\sfI)$, which is the desired result.

Statement~\ref{statement-1} will be immediate, and thus we shall focus on proving statement~\ref{statement-2} in Section~\ref{sec:insertion-BU}, and statement~\ref{statement-3} in Section~\ref{sec:deletion-BU}.

\subsection{Profiles and canonical rebalancing}

Let~$\scrT = (\calT^1,\calT^2,\ldots,\calT^\ell)$ be a cut of a tree~$\calT$ with root~$t$.
Denoting each rank difference~$\sfr(\calT) - \sfr(\calT^i)$ by~$\Delta\sfr^i$, the \emph{ordered profile} and the \emph{profile} of~$\scrT$ are defined as the list~$(\Delta\sfr^1,\Delta\sfr^2,\ldots,\Delta\sfr^\ell)$ and as the multiset~$\{\Delta\sfr^1,\Delta\sfr^2,\ldots,\Delta\sfr^\ell\}$.
We abusively say that these are also (ordered) profiles of the tree~$\calT$ and of its root~$t$.
Hence, each tree~$\calT$ has profile~$\{0\}$, as witnessed by the cut containing~$\calT$ itself.

A prominent example of cut is, for each integer~$\delta \geqslant 0$, the \emph{$\delta$-cut} of~$\calT$, which consists of those maximal sub-trees~$\calT^1,\calT^2,\ldots,\calT^\ell$ of~$\calT$ for which~$\Delta\sfr^i \geqslant \delta$;
equivalently, the ancestors of this cut are the nodes~$a^i$ of~$\calT$ such that~$\sfr(a^i) \geqslant \sfr(\calT) - (\delta-1)$.

Finally, some~$\delta$-cuts with a specific profile come with \emph{canonical rebalancing} operations.
Let us see five such operations, numbered from~$\sfC\sfR_1$ to~$\sfC\sfR_5$, and illustrated in Figure~\ref{fig:3-cut}.
In each case, we describe how to rebalance the cut of a given tree with rank~$\sfr$;
other canonical rebalancing operations will be introduced in the next sections:

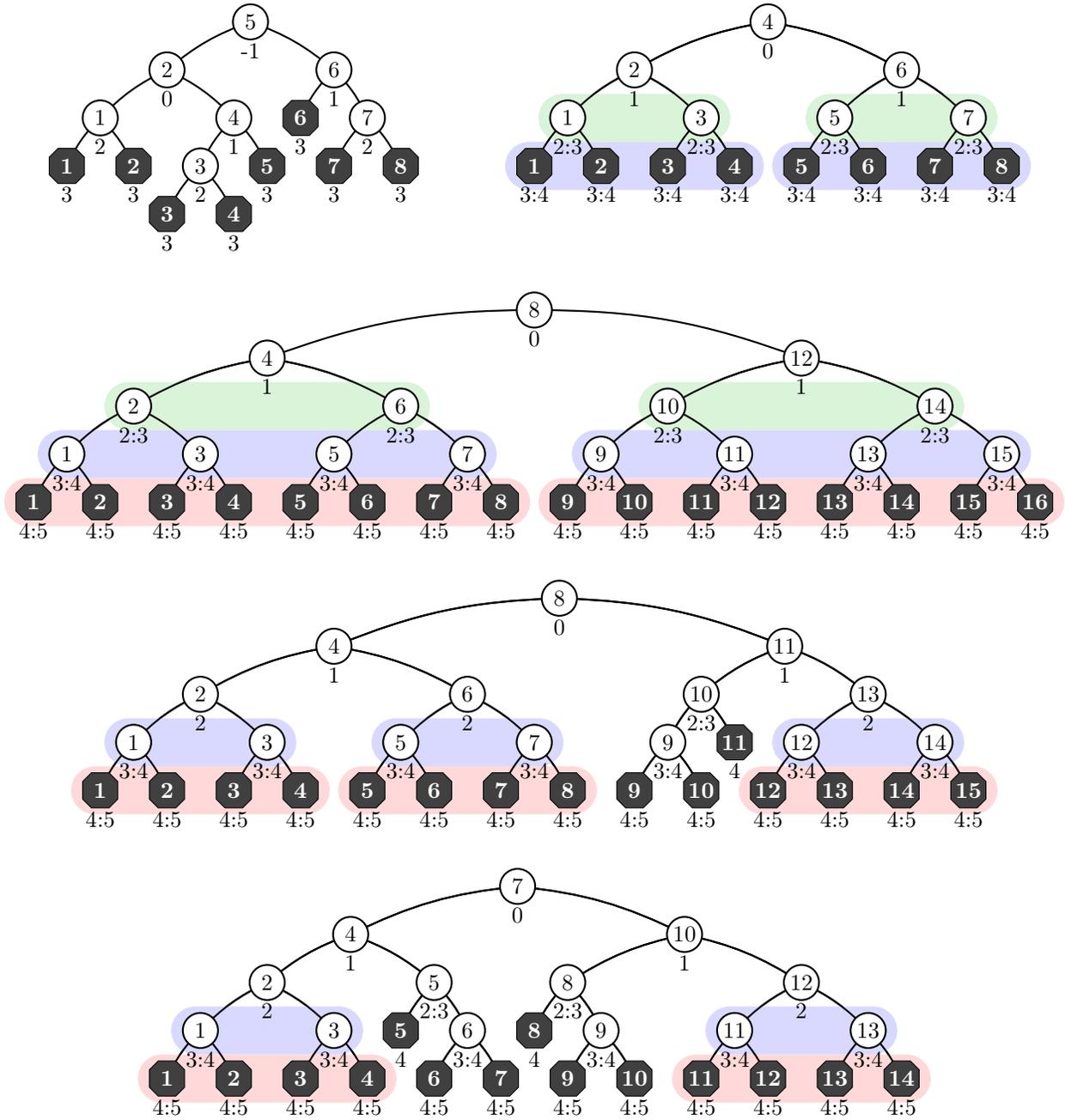
\begin{figure}[p]
\begin{center}
\begin{tikzpicture}[scale=0.8]
\renewcommand{\xlen}{0.646}
\renewcommand{\ylen}{0.93}
\place{1}{5.5/,3/1,1/11,0/111,2/112,5/12,4/121,3/1211,5/1212,6/122,%
8/2,7/21,9/22,8/221,10/222}
\labels{na}{11/2,1/0,121/2,12/1,/-1,2/1,22/2}
\labels{nb}{111/3,112/3,1211/3,1212/3,122/3,21/3,221/3,222/3}

\place{15}{7/,3/1,1/11,0/111,2/112,5/12,4/121,6/122,%
11/2,9/21,8/211,10/212,13/22,12/221,14/222}
\colorzone{11}{12}{green!70!black}
\colorzone{21}{22}{green!70!black}
\colorzone{111}{122}{blue}
\colorzone{211}{222}{blue}
\shape{,1,2,11,12,21,22}
\labels{na}{11/2:3,1/1,12/2:3,/0,21/2:3,2/1,22/2:3}
\labels{nb}{111/3:4,112/3:4,121/3:4,122/3:4,211/3:4,212/3:4,221/3:4,222/3:4}

\begin{scope}[shift={(0,-6*\ylen)}]
\place{0}{15/,7/1,23/2,3/11,11/12,19/21,27/22,%
1/111,5/112,9/121,13/122,17/211,21/212,25/221,29/222,%
0/1111,2/1112,4/1121,6/1122,%
8/1211,10/1212,12/1221,14/1222,%
16/2111,18/2112,20/2121,22/2122,%
24/2211,26/2212,28/2221,30/2222}
\colorzone{11}{12}{green!70!black}
\colorzone{21}{22}{green!70!black}
\colorzone{111}{122}{blue}
\colorzone{211}{222}{blue}
\colorzone{1111}{1222}{red}
\colorzone{2111}{2222}{red}
\shape{1,2,11,12,21,22,111,112,121,122,211,212,221,222}
\labels{na}{111/3:4,11/2:3,112/3:4,1/1,121/3:4,12/2:3,122/3:4,/0,%
211/3:4,21/2:3,212/3:4,2/1,221/3:4,22/2:3,222/3:4}
\labels{nb}{1111/4:5,1112/4:5,1121/4:5,1122/4:5,1211/4:5,1212/4:5,1221/4:5,1222/4:5,2111/4:5,2112/4:5,2121/4:5,2122/4:5,2211/4:5,2212/4:5,2221/4:5,2222/4:5}
\end{scope}

\begin{scope}[shift={(0,-12*\ylen)}]
\place{2}{13.75/,7/1,20.5/2,3/11,11/12,18/21,23/22,%
1/111,5/112,9/121,13/122,17/211,19/212,21/221,25/222,%
0/1111,2/1112,4/1121,6/1122,8/1211,10/1212,12/1221,14/1222,%
16/2111,18/2112,20/2211,22/2212,24/2221,26/2222}
\colorzone{111}{112}{blue}
\colorzone{121}{122}{blue}
\colorzone{221}{222}{blue}
\colorzone{1111}{1122}{red}
\colorzone{1211}{1222}{red}
\colorzone{2211}{2222}{red}
\shape{,1,2,11,12,21,22,111,112,121,122,211,221,222}
\labels{na}{111/3:4,11/2,112/3:4,1/1,121/3:4,12/2,122/3:4,/0,%
211/3:4,21/2:3,2/1,221/3:4,22/2,222/3:4}
\labels{nb}{1111/4:5,1112/4:5,1121/4:5,1122/4:5,1211/4:5,1212/4:5,1221/4:5,1222/4:5,2111/4:5,2112/4:5,212/4,2211/4:5,2212/4:5,2221/4:5,2222/4:5}
\end{scope}

\begin{scope}[shift={(0,-18*\ylen)}]
\place{4}{10.5/,5.5/1,15.5/2,3/11,8/12,12/21,19/22,%
1/111,5/112,7/121,9/122,11/211,13/212,17/221,21/222,%
0/1111,2/1112,4/1121,6/1122,8/1221,10/1222,%
12/2121,14/2122,16/2211,18/2212,20/2221,22/2222}
\colorzone{111}{112}{blue}
\colorzone{221}{222}{blue}
\colorzone{1111}{1122}{red}
\colorzone{2211}{2222}{red}
\shape{,1,2,11,12,21,22,111,112,122,212,221,222}
\labels{na}{111/3:4,11/2,112/3:4,1/1,12/2:3,122/3:4,/0,%
21/2:3,212/3:4,2/1,221/3:4,22/2,222/3:4}
\labels{nb}{1111/4:5,1112/4:5,1121/4:5,1122/4:5,121/4,1221/4:5,1222/4:5,211/4,2121/4:5,2122/4:5,2211/4:5,2212/4:5,2221/4:5,2222/4:5}
\end{scope}
\end{tikzpicture}
\vspace{-0.8em}
\end{center}
\caption{Applying operations~$\sfC\sfR_1$ (row~1, left),~$\sfC\sfR_2$ (row~1, right),~$\sfC\sfR_3$ (row~2),~$\sfC\sfR_4$ (row~4) and~$\sfC\sfR_5$ (row~5) on~$3$- and~$4$-cuts of a tree of rank~$\sfr$.
Each green (resp., blue, red) area contains at least one node of rank~$\sfr-2$ (resp.,~$\sfr-3$,~$\sfr-4$).
\label{fig:3-cut}}
\end{figure}

\begin{enumerate}[itemsep=0.5pt,topsep=0.5pt,label=$\sfC\sfR_\arabic*$:]
\item When~$\calT$ is has profile~$\{3,3,3,3,3,3,3,3\}$, we reshape its~$3$-cut and the ancestors of this cut according to Figure~\ref{fig:3-cut} (row~1, left).
This yields a tree~$\calT'$ of rank~$\sfr+1$ whose root is~$a^5$.

\item When~$\calT$ has profile~$\{2,2,2,3,3\}$, the~$3$-cut of~$\calT$ consists of eight trees, no three consecutive trees having rank~$\sfr-4$.
We reshape this~$3$-cut as a complete binary tree~$\calT'$, according to Figure~\ref{fig:3-cut} (row~1, right).
Since at least one of the trees~$\calT^1$ to~$\calT^4$ has rank~$\sfr-3$, the node~$a^2$ has rank~$\sfr-1$;
so does~$a^6$, and the root~$a^4$ of~$\calT'$ is a~$\{1,1\}$-node with rank~$\sfr$.

\item When~$\calT$ has profile~$\{3,3,3,3,3,3,3,4,4\}$, the~$4$-cut of~$\calT$ consists of sixteen trees, nine of which (at least) have rank~$\sfr-4$.
We reshape this~$4$-cut as a complete binary tree~$\calT'$, according to Figure~\ref{fig:3-cut} (row~2).
Since at least one of the trees~$\calT^1$ to~$\calT^8$ has rank~$\sfr-4$, their ancestor~$a^4$ has rank~$\sfr-1$;
so does~$a^{12}$ and the root~$a^8$ of~$\calT'$ is a~$\{1,1\}$-node with rank~$\sfr$.

\item When~$\calT$ has profile~$\{3,3,3,3,3,3,4,4,4\}$, the~$4$-cut of~$\calT$ consists of fifteen trees, with at least nine trees of rank~$\sfr-4$, and no three consecutive trees of rank~$\sfr-5$.
Thus, one of the eight trees~$\calT^{2i+1}$ in odd position has rank~$\sfr-4$; let~$\calT^{2j+1}$ be such a tree.

We define~$a^{2i+1}$ as the parent of~$\calT^{2i+1}$ and~$\calT^{2i+2}$ when~$i \leqslant j-1$, and~$a^{2i}$ as the parent of~$\calT^{2i}$ and~$\calT^{2i+1}$ when~$i \geqslant j+1$, thereby obtaining a cut consisting of eight trees of height~$\sfh-3$ or~$\sfh-4$.
We reshape this cut as a complete binary tree~$\calT'$, as illustrated in Figure~\ref{fig:3-cut} (row~3) in case~$j = 5$, whose root we call~$x'$.

Each child~$x'_u$ of~$x'$ has itself a child, say~$x'_{uv}$, that is not an ancestor of~$\calT^{j+1}$.
Hence, the grand-children of~$x'_{uv}$ are four consecutive trees~$\calT^{2i+1}$, and one of them is of rank~$\sfr-4$, which means that~$x'_{uv}$ and~$x'_u$ have ranks~$\sfr-2$ and~$\sfr-1$.
Consequently,~$\calT'$ is rooted at a~$\{1,1\}$-node with rank~$\sfr$.

\item When~$\calT$ has profile~$\{3,3,3,3,3,4,4,4,4\}$, our strategy is similar, but more complicated.
The~$4$-cut of~$\calT$ consists of fourteen trees, with at least nine trees of rank~$\sfr-4$, and no three consecutive trees having rank~$\sfr-5$.
One of the seven trees~$\calT^{2i+1}$ in odd position must have rank~$\sfr-4$; let~$\calT^{2j+1}$ be the leftmost such tree.
Similarly, one of the seven trees~$\calT^{2i}$ in even position must have rank~$\sfr-4$; let~$\calT^{2k}$ be the rightmost such tree.

All~$j$ trees~$\calT^1,\calT^3,\ldots,\calT^{2j-1}$ and all~$7-k$ trees~$\calT^{2k+2},\calT^{2k+4},\ldots,\calT^{14}$ have rank~$\sfr-5$, which means that~$j+7-k \leqslant 5$, i.e., that~$\calT^{2j+1}$ lies left of~$\calT^{2k}$.

Hence, we define~$a^{2i+1}$ as the parent of~$\calT^{2i+1}$ and~$\calT^{2i+2}$ when~$i \leqslant j-1$ or~$i \geqslant 2k$, and~$a^{2i}$ as the parent of~$\calT^{2i}$ and~$\calT^{2i+1}$ when~$j+1 \leqslant i \leqslant k-1$, thereby obtaining a cut consisting of eight trees of height~$\sfh-3$ or~$\sfh-4$.
We reshape this cut as a complete binary tree~$\calT'$, as illustrated in Figure~\ref{fig:3-cut} (row~3) in case~$j = 2$ and~$k = 4$, whose root we call~$x'$.

All grand-children of~$x'$ have rank~$\sfr-3$ or~$\sfr-2$.
At least two of these are not the parents of~$\calT^{2j-1}$ nor~$\calT^{2k}$: their grand-children are four consecutive trees~$\calT^i$, and their rank is~$\sfr-2$.
Hence, either both children of~$x'$ have rank~$\sfr-1$, and~$x'$ is a~$\{1,1\}$-node, or one of them has rank~$\sfr-2$, which means that its children have ranks~$\sfr-3$ and that~$x'$ has profile~$\{2,2,2\}$.
In conclusion,~$\calT'$ is rooted at node of rank~$\sfr$ that has profile~$\{1,1\}$ or~$\{2,2,2\}$.
\end{enumerate}

Finally, we allow ourselves to apply these canonical rebalancing operations to cuts with \emph{shifted} profiles (e.g., by taking one of the above profiles and adding~$1$ to each of them).
For instance, when~$\calT$ has profile~$\{3,3,3,4,4\}$ instead of~$\{2,2,2,3,3\}$, we may still decide to apply~$\sfC\sfR_2$, which will yield a tree whose root is a~$\{1,1\}$-node of rank~$\sfr(\calT)-1$.

\subsection{Full node and full elimination}

Given an integer~$\delta \geqslant 0$, we say that a node~$x$ is~$\delta$-\emph{full} when~$\delta = 0$, or~$x$ is a~$\{0,1\}$-node or~$\{1,1\}$-node whose~$1$-children are~$(\delta-1)$-full; in particular, leaves are~$1$-full but null nodes are not.

Then, we say that a tree~$\calT$ is \emph{overfull} if its root~$x$ is a~$\{1,1\}$-node and if at least one of~$x$,~$x_1$ and~$x_2$ is a~$3$-full node.
Below, we will occasionally eliminate overfull trees, by using an operation called \emph{full elimination} and denoted by~$\sfF\sfE$.
This operation is a special case of cut rebalancing, and works as follows.
Without loss of generality, we assume that if~$x_2$ is~$3$-full, so is~$x_1$ (and, therefore, so is~$x$):

\begin{itemize}[leftmargin=4.8mm,itemsep=0.5pt,topsep=0.5pt]
\item If~$x$ is~$3$-full, we apply~$\sfC\sfR_1$ to the~$3$-cut of~$\calT$.
Denoting by~$a^1,\ldots,a^7$ the ancestors of this cut, and as illustrated in Figure~\ref{fig:3-cut} (row~1, left), the nodes whose induced sub-trees change are all transformed into~$\{1,2\}$-nodes;
they include~$a^2$,~$a^5$ and~$a^6$, which are the root of the resulting tree and its children.%
\label{case:fe1}

\item If~$x$ is not~$3$-full but~$x_1$ is~$3$-full, we promote~$x$ and apply~$\sfC\sfR_1$ to the~$3$-cut of~$\calT(x_1)$.
This transforms~$\calT(x_1)$ into a tree~$\calT'$ of rank~$\sfr$ whose root and children are not~$3$-full, and~$x$ itself into a~$\{1,2\}$-node of rank~$\sfr+1$.%
\label{case:fe2}
\end{itemize}

\subsection{Insertion}
\label{sec:insertion-BU}

As announced in Section~\ref{sec:overview-BU}, we wish to transform as many transient operations, which occur in cases~\ref{sota:case-1} and~\ref{sota:case-4} of the textbook algorithm, into terminal operations.

First, an immediate induction proves that the lower end~$x_i$ of a zero-edge of rank~$\sfr \geqslant 1$ has profile~$\{1,2\}$.
This proves the well-known fact that case~\ref{sota:case-1} is actually spurious.
Going further also proves that~$x_i$ has profiles~$\{2,2,3\}$ if~$\sfr \geqslant 2$, and~$\{2,3,3,4\}$ if~$\sfr \geqslant 3$.

With the help these remarks, we subdivide  case~\ref{sota:case-4} into three sub-cases.
More precisely, when aiming to eliminate or propagate upward a zero-edge of rank~$\sfr \geqslant 4$ between a node~$x$ and its left child~$x_1$, and provided that~$\sfr(x_2) = \sfr - 1$:

\begin{enumerate}[leftmargin=4.8mm,itemsep=0.5pt,topsep=0.5pt]
\setcounter{enumi}{3}
\item
\begin{enumerate}[leftmargin=4.8mm,itemsep=0.5pt,topsep=0.5pt,label=\alph*)]
\item If~$x_2$ is not a~$1$-full node, i.e., is a~$\{1,2\}$-node, recalling that~$x_1$ has profile~$\{2,2,3\}$ proves that~$x$ has profile~$\{2,2,2,3,3\}$.
Hence, applying~$\sfC\sfR_2$ transforms~$\calT(x)$ into a tree of rank~$\sfr$ from which the zero-edge was eliminated.%
\label{bu-case-4a}

\item If~$x_2$ is a~$1$-full node but not a~$2$-full node, it has profile~$\{2,2,3,3\}$ or~$\{2,2,2,3\}$.
Then, since~$x_1$ has profile~$\{2,3,3,4\}$, it has profile~$\{3,3,3,4,4\}$ or~$\{3,3,3,3,4\}$.
Consequently,~$x$ itself has profile~$\{3,3,3,3,3,3 \text{ or } 4, 3 \text{ or } 4, 4, 4\}$.
Hence, applying~$\sfC\sfR_3$,~$\sfC\sfR_4$ or~$\sfC\sfR_5$ transforms~$\calT(x)$ into a tree of rank~$\sfr$ from which the zero-edge was eliminated.%
\label{bu-case-4b}

\item If~$x_2$ is a~$2$-full node, we simply rebalance the cut~$\scrT = \calT(x_1,x_2)$: this is the same promotion as in the textbook algorithm.%
\label{bu-case-4c}
\end{enumerate}
\end{enumerate}

As a result, the only transient case is case~\ref{sota:case-4}\ref{bu-case-4c}, on the condition that~$x$ was a~$1$-child.
When this happens, the node~$x$, which was~$3$-full before being promoted, becomes a~$\{1,2\}$-node, and promoting~$x$ does not help any of its ancestors to become~$1$-,~$2$- or~$3$-full.

\subsection{Deletion}
\label{sec:deletion-BU}

Similarly, we wish to transform as many transient operations, which occur in cases~\ref{sota:case-6} to~\ref{sota:case-8} of the textbook algorithm, into terminal operations.

The rotations and demotion performed in these cases transform a tree~$\calT(x)$ of rank~$\sfr$ rooted at a four-node~$x$ into a tree~$\calT'$ of rank~$\sfr-1$ whose root~$x'$ is a~$\{1,1\}$-node.
Hence, if~$x'$ or one of its children is~$3$-full,~$\calT'$ is an overfull tree, and the~$\sfF\sfE$ operation will transform it into another tree~$\calT''$ of rank~$\sfr$, making the two-step transformation of~$\calT(x)$ into~$\calT''$ a terminal operation.

If, however, neither~$x'$ nor its children is~$3$-full, the rotation or demotion we performed in case~\ref{sota:case-6},~\ref{sota:case-7} or~\ref{sota:case-8} did not create any~$3$-full node.
Indeed, replacing~$\calT(x)$ by a tree of smaller rank did not make any ancestor of~$x$ a~$3$-full node.
The only other nodes whose rank or induced sub-tree was changed are~$x'$ and its children, which we precisely assumed are not~$3$-full either.

In that case, our operation is transient if the parent of~$x$, say~$y$, was a~$\{1,2\}$-node, and became a four-node.
Consequently:
(i)~in case~\ref{sota:case-6}, we destroyed two~$\{1,2\}$-nodes ($a^2$ and~$y$);
(ii)~in case~\ref{sota:case-7}, we destroyed two~$\{1,2\}$-nodes ($a^1$ and~$y$), possibly making one of the nodes~$a^1$ or~$a^3$ a~$\{1,2\}$-node in case~$a^2$ was also a~$\{1,2\}$-node we just destroyed;
(iii)~in case~\ref{sota:case-8}, we destroyed one~$\{1,2\}$-node ($y$).
Consequently, as desired, statement~\ref{statement-3} of Section~\ref{sec:overview-BU} is valid.

\section{Top-down updates}
\label{sec:td}

\subsection{Overview}
\label{sec:overview-td}

Both algorithms for inserting a key~$k$ in an AVL tree~$\calT$ or deleting~$k$ from~$\calT$ follow the same general ideas.
As usual, we will discover, in a top-down manner, the branch~$x^0,x^1,\ldots$ below which~$k$ should be inserted or deleted, where~$x^0$ is the root of~$\calT$ and each node~$x^{i+1}$ is a child of the previous node~$x^i$.
Using a~$\calO(1)$ look-ahead, we shall first discover the first~$\calO(1)$ nodes on that branch, and already decide what to do.

If we are lucky enough, we found a node~$x^i$, for some integer~$i \geqslant 1$, that is deemed \emph{safe}: propagating a zero-edge (during insertions) or~four-node (during deletions) that we would have planted along our branch far enough below~$x^{i+5}$ will stop when meeting the node~$x^i$.
When such a node~$x^i$ exists among the top few nodes of the branch, we say that~$\calT$ was \emph{friendly};
indeed, we should simply focus on updating the tree~$\calT(x^i)$, whose rank is smaller than~$\calT$ itself, and which will be transformed into a new tree of the same rank.

This fortunate case suggests focusing only on updating trees whose root is safe.
Hence, if~$\calT$ has an unsafe root, a preliminary step will consist in transforming~$\calT$ into a tree whose root is safe, which we will then update;
all subsequent (recursively called) updates will be performed on trees whose root is safe.

If, however,~$\calT$ has a safe root but is unfriendly, we shall choose a deep enough node, say~$x^\ell$, and transform the sub-tree~$\calT(x^\ell)$ into a tree~$\calT'$ whose root is safe.
This transformation may create a zero-edge (during insertions) or a~four-node (during deletions), which will be propagated upward, but whose propagation will stop, at the latest, when meeting the safe root~$x$;
it just remains to update~$\calT'$.

\subsection{Insertion}
\label{sec:insertion-td}

Below, we say that a node~$x$ is \emph{insertion-safe} (or \isafe) when~(i)~$\sfr(x) = 1$ and~$x$ is a~$\{1,2\}$-node, or (ii)~$\sfr(x) \geqslant 2$ and~$x$ is not~$3$-full.
We will prove by induction on~$\sfr$ that, provided the root~$x$ of~$\calT(x)$ is \isafe, inserting~$k$ into~$\calT$ transforms~$\calT$ into a new tree of rank~$\sfr$.

First, if~$\sfr \leqslant 6$, we simply use the bottom-up algorithm of Section~\ref{sec:insertion-BU}:
if~$\calT(x)$ is replaced by a tree of larger rank, the last write operation we performed consisted in promoting~$x$, which meant that~$x$ was~$3$-full at that time.
Since transient operations do not create any \iunsafe nodes, this means that~$x$ was \iunsafe from the beginning.

Second, if~$\calT(x)$ is rooted at an \iunsafe node~$x$ of rank~$\sfr \geqslant 7$, it must be overfull:
using~$\sfF\sfE$, we replace~$\calT(x)$ by a tree~$\calT'$ whose root is a~$\{1,2\}$-node, which makes it \isafe.

It remains to treat the case of a tree~$\calT(x)$ rooted at an \isafe node~$x$ of rank~$\sfr \geqslant 7$.
Let~$x^0,x^1,\ldots$ be the first nodes of the branch below which~$k$ shall be inserted,~$x^0$ being the node~$x$, and each node~$x^{i+1}$ being the child of~$x^i$;
we stop at the first node~$x^i$ that is not a~$\{1,1\}$-node, or at the node~$x^5$.

If one of these nodes (except~$x^0$ itself), say~$x^i$, is \isafe, we just focus on inserting~$k$ in~$\calT(x^i)$: the tree~$\calT(x)$ is \emph{insertion-friendly}.

Otherwise, since the node~$x^5$ has rank~$\sfr-5 \geqslant 2$, it must be~$3$-full.
Hence, we apply~$\sfF\sfE$ to~$\calT(x^5)$, which transforms it into a friendly tree~$\calT'$ of rank~$\sfr(\calT') = \sfr-4$;
this creates a zero-edge between~$x^4$ and the root~$x'$ or~$\calT'$.
We move that zero-edge upward by promoting~$x^3$,~$x^2$ and~$x^1$.
Then, if~$x^1$ was a~$1$-child of~$x^0$ (in~$\calT$), we perform the (terminal) operation dictated by cases~\ref{sota:case-2},~\ref{sota:case-3},~\ref{sota:case-4}.\ref{bu-case-4a} or~\ref{sota:case-4}.\ref{bu-case-4b} of the efficient bottom-up insertion algorithm, which eliminated the zero-edge;
if~$x^1$ was a~$2$-child of~$x^0$, the zero-edge is already eliminated.
Finally, it remains to insert~$k$ into~$\calT'$.

\subsection{Deletion}
\label{sec:deletion-td}

Here, unlike in Section~\ref{sec:insertion-TD}, we temporarily forget the improvements of our efficient bottom-up deletion algorithm, and focus only on its textbook variant.
In this context, a node~$x$ is \emph{deletion-safe} (of \dsafe) if~$\sfr(x) \geqslant 1$ and~$x$ has profile~$\{1,1\}$ or~$\{2,2,2\}$.
Indeed, if propagating four-nodes ever forces us to replace~$\calT(x)$ by a tree of smaller rank, we must previously have replaced some sub-tree, say~$\calT(x_1)$, by a tree~$\calT'$ of rank~$\sfr(\calT') = \sfr(x_1) - 1$:
(i)~if~$x$ used to be a~$\{1,1\}$-node, we just transformed it into a~$\{1,2\}$-node, and (ii)~if~$x$ used to be a~$\{2,2,2\}$-node, the case~(i) shows that~$x_1$ could not be a~$\{1,1\}$-node itself, and therefore we were in case~\ref{sota:case-5} of the textbook deletion algorithm.

We will then prove by induction on~$\sfr$ that, provided the root~$x$ of a tree~$\calT$ with rank~$\sfr$ is \dsafe, deleting~$k$ from~$\calT$ transforms~$\calT$ into a new tree of rank~$\sfr$.
In order to do so, we use two additional canonical rebalancing operations, numbered~$\sfC\sfR_6$ and~$\sfC\sfR_7$.
In each case, we describe how to rebalance the cut of a given tree with rank~$\sfr$:

\begin{figure}[t]
\begin{center}
\begin{tikzpicture}[scale=0.8]
\renewcommand{\xlen}{0.646}
\renewcommand{\ylen}{0.93}
\place{0}{7/,4/1,1/11,0/111,2/112,1/1121,3/1122,%
7/12,6/121,8/122,5/1211,7/1212,%
10/2,9/21,11/22,10/221,12/222}
\labels{na}{11/2,112/3,1/1,121/3,12/2,/0,2/2,22/3}
\labels{nb}{111/4,1121/4,1122/4,1211/5,1212/4,122/4,21/4,221/4,222/4}
\end{tikzpicture}
\vspace{-0.8em}
\end{center}
\caption{Applying operation~$\sfC\sfR_6$ to the~$4$-cut of a tree of rank~$\sfr$, in case~$\sfs(0) = \sfs(2) = 1$ and~$\sfs(1) = 2$.
\label{fig:cr6}}
\end{figure}
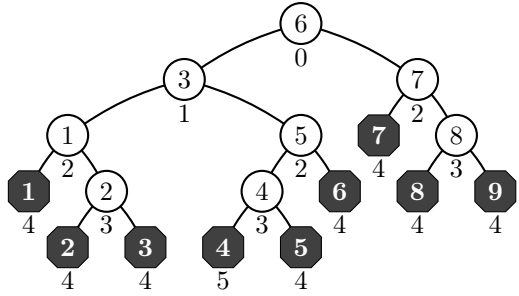

\begin{enumerate}[itemsep=0.5pt,topsep=0.5pt,label=$\sfC\sfR_\arabic*$:]
\setcounter{enumi}{5}

\item When~$\calT$ has profile~$\{4,4,4,4,4,4,4,4,4 \text{ or } 5\}$, we split the trees~$\calT^i$ in groups of three: we gather trees~$\calT^{3i+1}$,~$\calT^{3i+2}$ and~$\calT^{3i+3}$.
In each group, if~$\Delta\sfr^{3i+1} = 5$, we let~$a^{3i+1}$ and~$a^{3i+2}$ be the parents of~$\calT^{3i+1}$ and~$\calT^{3i+2}$, and~$a^{3i+1}$ and~$\calT^{3i+3}$, respectively;
in that case, we set~$\sfs(i) = 2$.
Otherwise, we let~$a^{3i+1}$ and~$a^{3i+2}$ be the parents of~$\calT^{3i+1}$ and~$a^{3i+2}$, and~$\calT^{3i+2}$ and~$\calT^{3i+3}$, respectively; in that case, we set~$\sfs(i) = 1$.

Doing so for each group of three trees provides us with nodes~$a^{3i+\sfs(i)}$ of rank~$\sfr-2$, which we group together by making~$a^3$ the parent of~$a^{\sfs(0)}$ and~$a^{3+\sfs(1)}$, and~$a^6$ the parent of~$a^3$ and~$a^{6+\sfs(2)}$.
This yields a tree of rank~$\sfr$ whose root has profile~$\{2,2,2\}$.
This process is illustrated in Figure~\ref{fig:cr6} in case~$\Delta\sfr^4 = 5$.

\item When the root~$x$ of~$\calT$ has rank~$\sfr \geqslant 4$ and has a child~$x_i$ such that neither~$x$ nor~$x_i$ are \dsafe,~$x$ is a~$\{1,2\}$-node whose~$1$-child is also a~$\{1,2\}$-node; hence,~$x$ has profile~$\{2,2,3\}$.
Furthermore, either its~$1$-child fails to have profile~$\{2,2,2\}$, or its~$2$-child fails to have profile~$\{1,1\}$, or both: in all cases,~$x$ cannot have profile~$\{3,3,3,3,3\}$.

Hence, the~$3$-cut of~$\calT$ has profile~$\{3,3,3,4,4\}$ or~$\{3,3,3,3,4\}$.
In the former case,~$\sfC\sfR_2$ transforms~$\calT$ into a tree of rank~$\sfr-1$ whose root is a~$\{1,1\}$-node.
In the latter case, the~$4$-cut of~$\calT$ has as profile~$\{4,4,4,4,4,4 \text{ or } 5,4 \text{ or } 5,4 \text{ or } 5,4 \text{ or } 5\}$, and therefore applying~$\sfC\sfR_3$, {}$\sfC\sfR_4$, {}$\sfC\sfR_5$ or~$\sfC\sfR_6$ transforms~$\calT(x)$ into a tree of rank~$\sfr-1$ or~$\sfr$ whose root has profile~$\{1,1\}$ or~$\{2,2,2\}$.
\end{enumerate}

Now, our algorithm works as follows.
First, if~$\sfr \leqslant 7$, we simply use the textbook bottom-up algorithm: we just proved that, if~$x$ is \dsafe, the tree~$\calT(x)$ will be replaced by a tree of the same rank.

Second, if~$\calT(x)$ is rooted at a node~$x$ of rank~$\sfr \geqslant 8$, let~$x^0 = x$ and let~$x^1$ be the child of~$x$ below which~$k$ shall be deleted.
If~$x^1$ is \dsafe, we just focus on deleting~$k$ from~$\calT(x^1)$.
Otherwise, if both~$x^0$ and~$x^1$ are \dunsafe, we apply~$\sfC\sfR_7$ to~$\calT(x)$, which transforms it into a tree whose root is \dsafe.

It remains to treat the case of a tree~$\calT(x)$ rooted at a \dsafe node~$x$ of rank~$\sfr \geqslant 8$.
Let~$x^0$,~$x^1$,~$x^2$ and~$x^3$ be the first nodes of the branch below which~$k$ shall be deleted,~$x^0$ being the node~$x$, and each node~$x^{i+1}$ being the child of~$x^i$.
If one of~$x^1$ or~$x^2$, say~$x^i$, is \dsafe, we just focus on deleting~$k$ in~$\calT(x^i)$: the tree~$\calT(x)$ is \emph{deletion-friendly}.

Otherwise, since the node~$x^2$ has rank~$\sfr(x^2) \geqslant \sfr-4 \geqslant 4$, we apply~$\sfF\sfE$ to~$\calT(x^2)$, which transforms it into a friendly tree~$\calT'$ of rank~$\sfr(\calT') = \sfr(x^2)$ or~$\sfr(x^2)-1$.
In the latter case, if~$x^1$ was a~$\{1,2\}$-node, it becomes a four-node, which will be propagated upward until it is eliminated;
at the latest, the elimination takes place just after the \dsafe node~$x^0$ has been transformed into a four-node.
Finally, it remains to delete~$k$ from~$\calT'$.

\section{Efficient top-down updates}
\label{sec:TD}

\subsection{Overview and complexity analysis}
\label{sec:overview-TD}

What made the success of the algorithm of Section~\ref{sec:BU} is that transient write operations never create~$3$-full nodes, and destroy either~$3$-full nodes or~$\{1,2\}$-nodes.
However, in the top-down algorithm of Section~\ref{sec:td}, these write operations come along with other transformations, aimed at making a node safe or a stopping the propagation of a zero-edge or four-node;
these other transformations may create both~$3$-full nodes and~$\{1,2\}$-nodes.

Consequently, we should make sure that, each time we make a node safe, we also perform a large number, say~$d$, of transient operations, and finally an operation that stops the propagation, and would have been considered terminal in Section~\ref{sec:BU}.
Indeed, taken together, these operations will either
(i)~during insertions, destroy~$d$, or slightly less,~$3$-full nodes, and (ii)~during deletions, create a constant number of~$3$-full nodes and destroy~$d$, or slightly less,~$\{1,2\}$-nodes.

More precisely, like in Section~\ref{sec:td}, we will discover, in a top-down manner, the branch~$x^0,x^1,\ldots$ below which~$k$ should be inserted or deleted, where~$x^0$ is the root of~$\calT$ and each node~$x^{i+1}$ is a child of the previous node~$x^i$.

If~$\calT$ has a large enough rank, but none of the first few nodes~$x^i$ is safe, we first make~$\calT$ a tree whose root is safe.
Hence, possibly after performing this preliminary step, we just focus on updating trees whose roots are safe.

Then, if we are lucky enough, among the~$\calO(1)$ first nodes~$x^1,x^2,\ldots$ on the branch we shall update, we found a safe node~$x^i$;
in that case,~$\calT$ is friendly, and we should simply focus on updating the sub-tree~$\calT(x^i)$.

If, however,~$\calT$ has a safe root but is unfriendly, we shall transform a deep enough sub-tree~$\calT(x^d)$ into a tree whose root~$\calT'$ is safe, but we shall also ensure that this creates a zero-edge (during insertions) or a four-node (during deletions), that will be propagated upward~$d$ times, until the safe node~$x^0$ stops this propagation.
It will then remain to update~$\calT'$.

Beside the preliminary step and low-height operations, which trigger a constant number of write operations per update,  all our write operations consist in a \emph{batch} of operations: making a transformation, followed by~$d$ transient operations, and finally one operation that would have been considered terminal if we were using the bottom-up algorithm.

Now, our complexity analysis works as follows.
Here and in the rest of the article, we set~$d = 58$, with the idea that each batch should include~$d$ transient operations.
Under those conditions, we will prove that exists a constant~$\sfK$ with the following properties (the explicit constants provided below are certainly not optimal, but simple enough to prove):

\begin{enumerate}[leftmargin=5.8mm,itemsep=0.5pt,topsep=0.5pt]
\item during insertions, each batch increases the total number of~$3$-full nodes by no more than~$27 - d$, and the total number of~$\{1,2\}$-nodes by no more than~$25 + d$;%
\label{statement-4}
\item during deletions, each batch increases the total number of~$3$-full nodes by~$11$, and the total number of~$\{1,2\}$-nodes by~$14 - d$;
\label{statement-5}
\item finally, each update may require performing the preliminary step and low-height operations, which increases the total numbers of~$3$-full nodes and of~$\{1,2\}$-nodes by~$\sfK$ per update.
\label{statement-6}
\end{enumerate}

If~$n$ updates trigger~$\sfI$ insertion batches and~$\sfD$ deletion batches, starting from an empty tree, the number of~$3$-full nodes increases by no more than~$(27 - d) \sfI + 11 \sfD+ \sfK n$, and the number of~$\{1,2\}$-nodes increases by~$(25+d) \sfI + (14 - d) \sfD + \sfK n$, two quantities that are thus bounded from below by~$0$.
It follows that~$(27 - d) \sfI + 11 \sfD+ \sfK n \geqslant 0$ and~$(25+d) \sfI + (14 - d) \sfD + \sfK n \geqslant 0$, two inequalities that can be rewritten, when~$d = 58$, as~$\sfI \leqslant 4 \sfK n/41$ and~$\sfD \leqslant 114 \sfK n / 451$.

In conclusion, these~$n$ updates trigger~$\calO(n)$ batches, preliminary steps and low-height operations, which account for a total of~$\calO(n)$ write operations.
Below, we present our insertion and deletion algorithms, and we will finally check the validity of statements~\ref{statement-4} and~\ref{statement-5}; statement~\ref{statement-6} will be immediate.

\subsection{Insertion}
\label{sec:insertion-TD}

The algorithm we use is very similar to that of Section~\ref{sec:insertion-td}.
The only differences are that
(i)~low-height cases concern trees of rank~$\sfr \leqslant d+2$, and not only~$\sfr \leqslant 6$;
(ii)~a tree~$\calT(x)$ is considered friendly if one of the first~$d+1$ nodes on the branch we shall update is \isafe, instead of the five first nodes only.

Hence, we just need to focus on what happens if an unfriendly tree~$\calT(x)$ is rooted at an \isafe node of rank~$\sfr \geqslant d+3$.
In that case, all nodes~$x^1,\ldots,x^{d+1}$ have rank at least~$\sfr - (d+1) \geqslant 2$, hence they are~$3$-full.
We apply~$\sfF\sfE$ to~$\calT(x^{d+1})$, which transforms it into a new tree~$\calT'$ of rank~$\sfr-d$, whose root~$x'$ is \isafe.
This creates a zero-edge between~$x^{d}$ and~$x'$, which we propagate upward by promoting all nodes~$x^{d},x^{d-1},\ldots,x^1$.
Finally, either~$x^1$ used to be a~$2$-child of~$x^0$ (in~$\calT$), in which case we are done, or we apply one of the transformations dictated by cases~\ref{sota:case-2} to~\ref{sota:case-4} of the efficient insertion algorithm.

Transforming~$\calT(x^{d+1})$ into~$\calT'$ and performing our last write operation affected the vicinity and rank of a constant number of nodes, thereby creating no more than a constant number of~$3$-full nodes and a constant number of~$\{1,2\}$-nodes.
In the meantime, each of the~$d$ promotions we performed destroyed a~$3$-full node.
Consequently, as desired, statement~\ref{statement-4} of Section~\ref{sec:overview-TD} is valid.

\subsection{Deletion}
\label{sec:deletion-TD}

The algorithm we use is less similar to that of Section~\ref{sec:deletion-td}, because we must change our notion of safe nodes.

A node~$x$ shall be considered \emph{efficient-deletion-safe}, or \edsafe when (i)~$\sfr(x) \leqslant 1$ and~$x$ is a~$\{1,1\}$-node, or when~$x$ has profile (ii)~$\{2,2,2\}$ or (iii)~$\{4,4,4,4,4,4,4,4 \text{ or } 5,4 \text{ or } 5\}$, or when~$\sfr(x) \geqslant 5$ and~$x$ has ordered profile (iv)~$(2,5,5,5,5,2)$ or (v)~$(1,4,4,4,4)$ or (vi)~$(3,3 \text{ or } 4,5,5,5,5,2)$ or (vii)~$(4,4,4 \text{ or } 5, 4 \text{ or }5,4,3 \text{ or } 4,2)$ or, symmetrically, (v)'~$(4,4,4,4,1)$ or (vi')~$(2,5,5,5,5,3 \text{ or } 4,3)$ or (vii')~$(2,3 \text{ or } 4,4,4 \text{ or } 5,4 \text{ or } 5,4,4)$.
In cases~(i) and~(ii), we still say that~$x$ is \dsafe.
When~$x$ is not \dsafe but falls in one of the cases~(iii) to~(vii'), we say that~$x$ is \emph{weakly} \edsafe: these are the cases that would trigger the creation of a~$3$-full node when using the textbook algorithm or, equivalently, stop the propagation of four-nodes when using the algorithm of Section~\ref{sec:deletion-BU}.

We will then prove by induction that, provided the root~$x$ of a tree~$\calT$ with rank~$\sfr$ is \edsafe, deleting~$k$ from~$\calT$ transforms~$\calT$ into a new tree of rank~$\sfr$.
This requires adapting the canonical rebalancing operations~$\sfC\sfR_6$ and~$\sfC\sfR_7$ introduced in Section~\ref{sec:deletion-td}, and proposing a new operation numbered~$\sfC\sfR_8$, which is a restriction of~$\sfC\sfR_7$:

\begin{enumerate}[itemsep=0.5pt,topsep=0.5pt,label=$\sfC\sfR_\arabic*$:]
\setcounter{enumi}{7}

\item When the root~$x$ of~$\calT$ has rank~$\sfr \geqslant 4$ and has a child~$x_i$ such that neither~$x$ nor~$x_i$ are \edsafe,~$\sfC\sfR_7$ transforms~$\calT$ into a tree of rank~$\sfr-1$ whose root is a~$\{1,1\}$-node.
Indeed, as indicated when we presented~$\sfC\sfR_7$, this happens when~$x$ has profile~$\{3,3,3,4,4\}$, where we would call~$\sfC\sfR_2$, or~$\{4,4,4,4,4,4 \text{ or } 5,4 \text{ or } 5,5,5\}$, where we would call~$\sfC\sfR_3$,~$\sfC\sfR_4$,~$\sfC\sfR_5$; but not when~$x$ has profile~$\{4,4,4,4,4,4,4,4,4 \text{ or } 5\}$, which would however mean that~$x$ is \edsafe.
\end{enumerate}

It also requires more subtle rebalancing operations, which we call \emph{node strengthening}; they are numbered from~$\sfN\sfS_3$ to~$\sfN\sfS_7$, thus matching the cases~(iii) to~(vii) that make a node weakly \edsafe, and illustrated in Figure~\ref{fig:strong-safe}.
Given a weakly \edsafe node~$x$ and a descendant~$y$ of~$x$ with rank~$\sfr(y) \leqslant \sfr(x) - 5$, the node strengthening rebalances the~$4$-cut of~$x$ to make one of the ancestors of this cut a \dsafe ancestor of~$y$.
In each case, we describe how to rebalance the cut of a given tree of rank~$\sfr$ whose root~$x$ is weakly \edsafe:

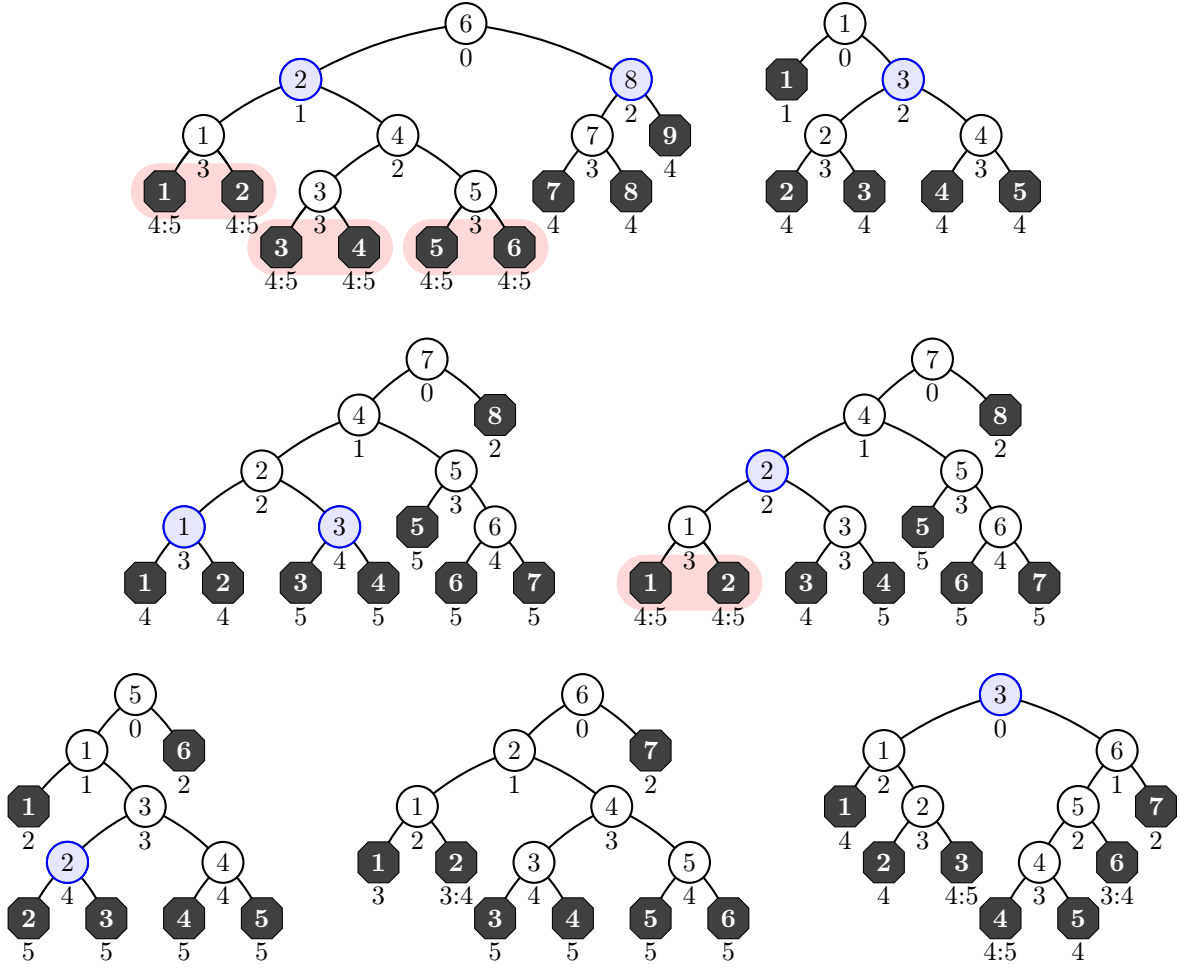
\begin{figure}[t]
\begin{center}
\begin{tikzpicture}[scale=0.8]
\renewcommand{\xlen}{0.646}
\renewcommand{\ylen}{0.93}
\place{3.5}{7.75/,3.5/1,1/11,0/111,2/112,6/12,4/121,3/1211,5/1212,%
8/122,7/1221,9/1222,12/2,11/21,10/211,12/212,13/22}
\colorzone{111}{112}{red}
\colorzone{1211}{1212}{red}
\colorzone{1221}{1222}{red}
\shape{11,121,122}
\labels{na}{11/3,1/1,121/3,12/2,122/3,/0,21/3,2/2}
\labels{nb}{111/4:5,112/4:5,1211/4:5,1212/4:5,1221/4:5,1222/4:5,%
211/4,212/4,22/4}
\labels{nd}{1/2,2/8}

\place{19.5}{1.5/,0/1,3/2,1/21,0/211,2/212,5/22,4/221,6/222}
\labels{na}{/0,21/3,2/2,22/3}
\labels{nb}{1/1,211/4,212/4,221/4,222/4}
\labels{nd}{2/3}

\begin{scope}[shift={(0*\xlen,-6*\ylen)}]
\place{3}{7.25/,5.5/1,3/11,1/111,5/112,8/12,7/121,9/122,%
0/1111,2/1112,4/1121,6/1122,8/1221,10/1222,9/2}
\labels{na}{111/3,11/2,112/4,1/1,12/3,122/4,/0}
\labels{nb}{1111/4,1112/4,1121/5,1122/5,121/5,1221/5,1222/5,2/2}
\labels{nd}{111/1,112/3}

\place{16}{7.25/,5.5/1,3/11,1/111,5/112,8/12,7/121,9/122,%
0/1111,2/1112,4/1121,6/1122,8/1221,10/1222,9/2}
\colorzone{1111}{1112}{red}
\shape{111}
\labels{na}{111/3,11/2,112/3,1/1,12/3,122/4,/0}
\labels{nb}{1111/4:5,1112/4:5,1121/4,1122/5,121/5,1221/5,1222/5,2/2}
\labels{nd}{11/2}
\end{scope}

\begin{scope}[shift={(0*\xlen,-12*\ylen)}]
\place{0}{2.75/,1.5/1,0/11,%
3/12,1/121,0/1211,2/1212,5/122,4/1221,6/1222,4/2}
\labels{na}{1/1,121/4,12/3,122/4,/0}
\labels{nb}{11/2,1211/5,1212/5,1221/5,1222/5,2/2}
\labels{nd}{121/2}

\place{9}{5.25/,3.5/1,1/11,0/111,2/112,%
6/12,4/121,3/1211,5/1212,8/122,7/1221,9/1222,7/2}
\labels{na}{11/2,1/1,121/4,12/3,122/4,/0}
\labels{nb}{111/3,112/3:4,1211/5,1212/5,1221/5,1222/5,2/2}

\place{21}{4/,1/1,0/11,2/12,1/121,3/122,%
7/2,6/21,5/211,4/2111,6/2112,7/212,8/22}
\labels{na}{1/2,12/3,/0,211/3,21/2,2/1}
\labels{nb}{11/4,121/4,122/4:5,2111/4:5,2112/4,212/3:4,22/2}
\labels{nd}{/3}
\end{scope}
\end{tikzpicture}
\vspace{-0.8em}
\end{center}
\caption{Making~$y$ descend from a (blue-painted) \dsafe node in case~(iii.c) (row~1, left); in case~(v.b) (row~1, right);
in case~(iv), when~$\sfr(\calU^3) = \sfr-5$ (row~2, left) or~$\sfr(\calU^3) = \sfr-4$ (row~2, right) or~$y$ descends from~$\calT^2$ or~$\calT^3$ (row~3, left);
in case~(vii) (row~3, right) ;
or transforming~$\calT$ into a tree with ordered profile~$(2,5,5,5,5,2)$ (row~3, centre) in case~(vi).
Each red area contains a node of rank~$\sfh-4$.
\label{fig:strong-safe}}
\end{figure}

\begin{enumerate}[itemsep=0.5pt,topsep=0.5pt,label=$\sfN\sfS_\arabic*$:]
\setcounter{enumi}{2}

\item When~$\calT$ has profile~$\{4,4,4,4,4,4,4,4,4 \text{ or } 5\}$, just use~$\sfC\sfR_6$: it transforms~$\calT$ into a tree of rank~$\sfr$ whose root is \dsafe.
When~$\calT$ has a cut~$\scrT$ with profile~$\{4,4,4,4,4,4,4,5,5\}$, however, let~$\calT^u$ and~$\calT^v$ be those trees with rank~$\sfr-5$;
we distinguish several cases:

\begin{enumerate}[itemsep=0.5pt,topsep=0.5pt]
\item If~$v = u+1$, defining~$a^u$ as their parent yields a cut~$\scrU$ consisting of eight trees with rank~$\sfr-4$, which we rebalance using~$\sfC\sfR_1$.
As a result, and denoting the ancestors of~$\scrU$ by~$b^1,\ldots,b^7$, the node~$y$ shall now descend from one of the \dsafe nodes~$b^1$,~$b^4$ or~$b^6$, as illustrated in Figure~\ref{fig:3-cut} (row~1, left).

\item If, by splitting the trees in groups of three, we put~$\calT^u$ and~$\calT^v$ in distinct groups, we follow the instructions of~$\sfC\sfR_6$, as illustrated in Figure~\ref{fig:cr6}:
each group is rebalanced to form a small tree of rank~$\sfr-2$, and these small trees are then organised in the shape of a tree whose root has profile~$\{2,2,2\}$.

\item If, on the contrary,~$\calT^u$ and~$\calT^v$ were put in the same group, we assume without loss of generality that~$1 \leqslant u \leqslant 6$, as illustrated in Figure~\ref{fig:strong-safe} (row~1, left), and~$y$ shall now descend from one of the~$\{2,2,2\}$-nodes~$a^2$ or~$a^8$.
\end{enumerate}

\item When~$\sfr \geqslant 5$ and~$\calT$ has a cut~$\scrT$ with ordered profile~$(2,5,5,5,5,2)$, we assume without loss of generality that~$y$ descends from one of the trees~$\calT^1$ to~$\calT^3$.

If~$y$ descends from~$\calT^1$, we may also assume that the root~$t^1$ of~$\calT^1$ is \dunsafe, which means it must have profile~$\{2,2,3\}$.
Hence, let~$\scrU = (\calU^1,\ldots,\calU^8)$ be the cut of~$\calT$ obtained when replacing~$\calT^1$ by its own~$2$-cut:
it has ordered profile~$(4,4,5,5,5,5,5,2)$ or~$(4,5,4,5,5,5,5,2)$ or~$(5,4,4,5,5,5,5,2)$:

\begin{enumerate}[itemsep=0.5pt,topsep=0.5pt]
\item if~$\calU^3$ has rank~$\sfr(x)-5$, and as illustrated in Figure~\ref{fig:strong-safe} (row~2, left),~$y$ shall now descend from the~$\{1,1\}$-node~$a^2$;
\item if~$\calU^3$ has rank~$\sfr(x)-4$, and as illustrated in Figure~\ref{fig:strong-safe} (row~2, right),~$y$ shall now descend from one of the~$\{1,1\}$-nodes~$a^1$ and~$a^3$.
\end{enumerate}

On the contrary,
\begin{enumerate}[itemsep=0.5pt,topsep=0.5pt]
\setcounter{enumii}{2}
\item if~$y$ descends from~$\calT^2$ or~$\calT^3$, and as illustrated in Figure~\ref{fig:strong-safe} (row~3, left), it shall now descend from the~$\{1,1\}$-node~$a^2$.
\end{enumerate}

\item When~$\calT$ has a cut~$\scrT$ with ordered profile~$(1,4,4,4,4)$:
\begin{enumerate}[itemsep=0.5pt,topsep=0.5pt]
\item if~$y$ descends from~$\calT^1$, we may assume that the root~$t^1$ of~$\calT^1$ is \dunsafe, which means it must have profile~$\{2,2,3\}$;
hence,~$\calT$ has profile~$\{3,3,4,4,4,4,4\}$, and its~$4$-cut has profile~$\{4,4,4,4,4,4,4,4 \text{ or } 5,4 \text{ or } 5\}$, which means that we can simply use~$\sfN\sfS_3$.

\item if~$y$ descends from some tree~$\calT^2$ to~$\calT^5$, and as illustrated in Figure~\ref{fig:strong-safe} (row~1, right), it shall now descend from the~$\{1,1\}$-node~$a^3$.
\end{enumerate}

\item When~$\sfr \geqslant 6$ and~$\calT$ has a cut~$\scrT$ with ordered profile~$(3,3 \text{ or } 4,5,5,5,5,2)$ or its symmetric, we proceed in two steps:
first, and as illustrated in Figure~\ref{fig:strong-safe} (row~3, centre), we rebalance~$\scrT$ to obtain a tree~$\calT'$ with ordered profile~$(2,5,5,5,5,2)$;
then, we use~$\sfN\sfS_3$ on~$\calT'$.

\item When~$\calT$ has a cut~$\scrT$ with ordered profile~$(4,4,4 \text{ or } 5,4 \text{ or } 5,4,3 \text{ or } 4,2)$ or its symmetric, and as illustrated in Figure~\ref{fig:strong-safe} (row~3, right), we rebalance~$\scrT$ to obtain a tree of rank~$\sfr$ whose root has profile~$\{2,2,2\}$.
\end{enumerate}

Now, our algorithm for deleting a key~$k$ from a tree~$\calT$ with rank~$\sfr$ works as follows.

First, if~$\sfr \leqslant 2d+18$ and the root~$x$ of~$\calT$ is weakly \edsafe, let~$y$ be the leaf of~$\calT$ containing the key~$k$;
by applying some operation~$\sfN\sfS_3$ to~$\sfN\sfS_6$ to~$\calT$, we ensure that~$y$ now descends from a \dsafe node~$x'$, and we just focus on deleting~$k$ from the sub-tree rooted at~$x'$.
If, however,~$x$ is either \dsafe or \edunsafe, we simply use the textbook bottom-up algorithm.

Second, if~$\calT(x)$ is rooted at a node~$x$ of rank~$\sfr \geqslant 2d+19$, let~$x^0,x^1,\ldots,x^{d+7}$ be the first nodes on the branch below which~$k$ should be deleted, where~$x^0 = x$ and each node~$x^{i+1}$ is a child of the previous node~$x^i$;
observing that each node~$x^i$ has rank~$\sfr(x^i) \geqslant \sfr - 2i$ proves that these nodes indeed exist and have rank~$5$ or more.
If one of these nodes other than~$x^0$, say~$x^j$ is \edsafe, we say that~$\calT(x)$ is friendly, and we just focus on deleting~$k$ from~$\calT(x^j)$.

If, however,~$\calT(x)$ is unfriendly and~$x$ is \edunsafe, and like in Section~\ref{sec:deletion-td}, applying~$\sfC\sfR_8$ to~$\calT(x)$ transforms it into a tree of rank~$\sfr$ whose root is \edsafe.
This case can occur only in a preliminary step.

Then, if~$\calT(x)$ is unfriendly and~$x$ is weakly \dsafe, we ensure by applying some operation~$\sfN\sfS_3$ to~$\sfN\sfS_6$ to~$\calT(x)$ that~$x^5$ now descends from a \dsafe node~$z$ of rank~$\sfr(z) \geqslant \sfr-4$;
this operation did not modify any sub-tree of rank~$\sfr-5$ or less.
Choosing~$z$ of minimal rank, the tree~$\calT'$ now rooted at~$z$ contains a branch~$z^0,z^1,\ldots$ below which the key~$k$ should be deleted.
This branch coincides with the major part of~$x^0,x^1,\ldots$:
indeed, denoting by~$u$ and~$v$ the least integers such that~$\sfr(x^u) \leqslant \sfr-5$ and~$\sfr(z^v) \leqslant \sfr-5$, we now have~$x^{u+i} = z^{v+i}$ for all~$i \geqslant 0$.
In particular, setting~$\delta = v-u \geqslant -5$, all nodes~$z^1,z^2,\ldots,z^{d+7+\delta}$ are \edunsafe.

As a result, we say that a tree~$\calT(x)$ is \emph{near-unfriendly} if all nodes~$x^1,x^2,\ldots,x^{d+2}$ are \edunsafe:
we reduced the study of unfriendly trees rooted at an \edsafe node to that of near-unfriendly trees rooted at a \dsafe node~$x$ of rank~$\sfr \geqslant 2d+15$.
In this latter case, since both~$x^{d+1}$ and~$x^{d+2}$ have rank at least~$11$ and are \edunsafe, we apply~$\sfC\sfR_8$ to~$\calT(x^{d+1})$, which transforms it into a new tree~$\calT'$ of rank~$\sfr(x^{d+1})-1$, whose root~$x'$ is~\edsafe.

The node~$x^{d}$, which was a~$\{1,2\}$-node, is thus transformed into a four-node, and we propagate this four-node upward by applying the rotations or demotions prescribed by the textbook algorithm.
Since no node~$x^i$ except~$x^0$ is \dsafe, the four-node is propagated until it is stopped by~$x^0$, which either becomes a~$\{1,2\}$-node or triggers the rotation of case~\ref{sota:case-5} of the textbook algorithm.
Hence, we performed~$d$ transient operations;
as noted in Section~\ref{sec:deletion-BU}, each of them decreases the number of~$\{1,2\}$-nodes, and it remains to investigate whether they can create~$3$-full nodes.

Given an integer~$i$ such that~$1 \leqslant i \leqslant d$, let us assume without loss of generality that~$x^{i+1}$ is the right child if~$x^i$, i.e., that~$x^{i+1} = x_1^i$.
When~$\calT(x^i)$ is replaced by a tree~$\calT'$ with root~$x'$ and rank~$\sfr(x^i)-1$, one of case~\ref{sota:case-6},~\ref{sota:case-7} or~\ref{sota:case-8} is triggered.
The only~$3$-full nodes it could create are those nodes of~$\calT(x^i)$ whose rank or induced sub-trees were modified.

By considering the three cases separately, and each of these nodes in each case, we can observe that creating a~$3$-full node would require~$x^i$ to be \edsafe in the first place (i.e., before~$\calT(x^{i+1})$ was replaced by a tree of smaller rank).
Each case is illustrated in Figure~\ref{fig:no-3-full}, where we represent the tree before~$\calT(x^{i+1})$ saw its height decrease, when~$x^i$ became a four-node, and after the four-node was propagated upward:

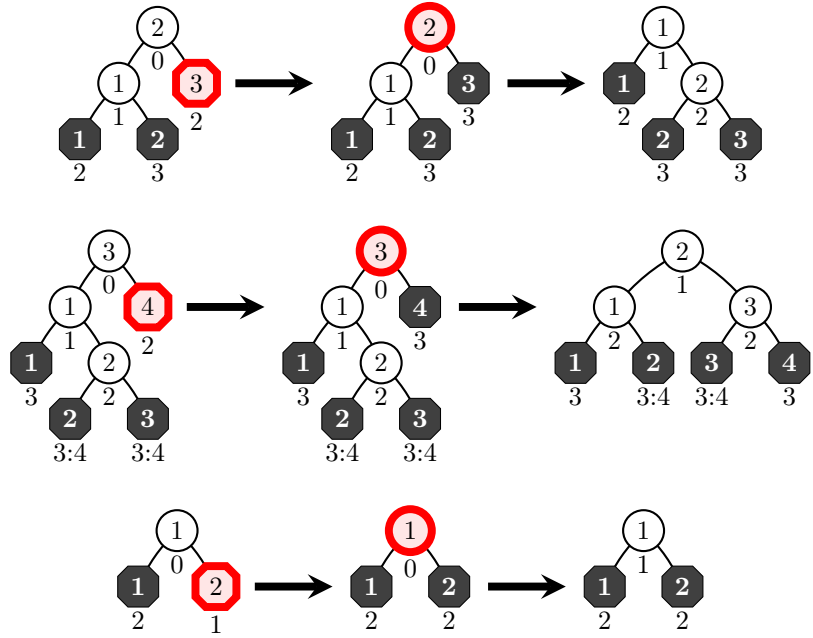
\begin{figure}[t]
\begin{center}
\begin{tikzpicture}[scale=0.8]
\renewcommand{\xlen}{0.646}
\renewcommand{\ylen}{0.93}
\begin{scope}[shift={(1.25*\xlen,0)}]
\place{0}{2/,1/1,0/11,2/12,3/2}
\labels{na}{1/1,/0}
\labels{nb}{11/2,12/3}
\nenode{2}{2}{3}

\arrow{4}

\place{7}{2/,1/1,0/11,2/12,3/2}
\labels{na}{1/1}
\ncnode{}{0}{2}
\labels{nb}{11/2,12/3,2/3}

\arrow{11}

\place{14}{1/,0/1,2/2,1/21,3/22}
\labels{na}{/1,2/2}
\labels{nb}{1/2,21/3,22/3}
\end{scope}


\begin{scope}[shift={(0,-4*\ylen)}]
\place{0}{2/,1/1,0/11,2/12,1/121,3/122,3/2}
\labels{na}{1/1,12/2,/0}
\labels{nb}{11/3,121/3:4,122/3:4}
\nenode{2}{2}{4}

\arrow{4}

\place{7}{2/,1/1,0/11,2/12,1/121,3/122,3/2}
\labels{na}{1/1,12/2}
\ncnode{}{0}{3}
\labels{nb}{11/3,121/3:4,122/3:4,2/3}

\arrow{11}

\place{14}{2.75/,1/1,0/11,2/12,4.5/2,3.5/21,5.5/22}
\labels{na}{1/2,/1,2/2}
\labels{nb}{11/3,12/3:4,21/3:4,22/3}
\end{scope}


\begin{scope}[shift={(2.75*\xlen,-9*\ylen)}]
\place{0}{1/,0/1,2/2}
\labels{na}{/0}
\labels{nb}{1/2}
\nenode{2}{1}{2}

\arrow{3}

\place{6}{1/,0/1,2/2}
\ncnode{}{0}{1}
\labels{nb}{1/2,2/2}

\arrow{9}

\place{12}{1/,0/1,2/2}
\labels{na}{/1}
\labels{nb}{1/2,2/2}
\end{scope}
\end{tikzpicture}
\vspace{-1.2em}
\end{center}
\caption{Decreasing the rank of~$\calT(x^{i+1})$ makes~$x^i$ a four-node, which is propagated upward without creating~$3$-full nodes, via the rotations of cases~\ref{sota:case-6} (row~1),~\ref{sota:case-7} (row~2) and~\ref{sota:case-8} (row~3).
\label{fig:no-3-full}}
\end{figure}

\begin{enumerate}
\setcounter{enumi}{5}

\item If the rotation turns~$a^1$ into a~$3$-full node, the sub-trees~$\calT^1$ and~$\calT^2$ must already have been~$2$-full and~$1$-full;
hence, before becoming a four-node,~$x^i$ had ordered profile~$(4,4,4,4,4,4,2)$, which is the case~(vii) for being \edsafe.

If this rotation turns~$a^2$ into a~$3$-full node, the sub-tree~$\calT^2$ must already have been~$2$-full;
hence, before becoming a four-node,~$x^i$ had ordered profile~$(2,5,5,5,5,2)$, which is the case~(iv) for being \edsafe.

\item If the rotation turns~$a^1$ into a~$3$-full node, the sub-trees~$\calT^1$ and~$\calT^2$ must already have been~$2$-full and have had rank~$\sfr-3$;
hence, before becoming a four-node,~$x^i$ had ordered profile~$(4,4,5,5,4,3 \text{ or } 4,2)$, which is the case~(vii) for being \edsafe.

If this rotation turns~$a^2$ into a~$3$-full node, the sub-trees~$\calT^1$,~$\calT^2$ and~$\calT^3$ must already have been~$1$-full and have had rank~$\sfr-3$;
hence, before becoming a four-node,~$x^i$ had ordered profile~$(4,4,4,4,4,4,2)$, which is the case~(vii) for being \edsafe.

Finally, if it turns~$a^3$ into a~$3$-full node, the sub-tree~$\calT^3$ must already have been~$2$-full and have had rank~$\sfr-3$;
hence, before becoming a four-node,~$x^i$ had ordered profile~$(3,3\text{ or } 4,5,5,5,5,2)$, which is the case~(vi) for being \edsafe.

\item If the demotion turns~$a^1$ into a~$3$-full node, the sub-tree~$\calT^1$ must already have been~$2$-full;
hence, before becoming a four-node,~$x^i$ had ordered profile~$(4,4,4,4,1)$, which is the case~(v') for being \edsafe.
\end{enumerate}

\subsection{Obtaining actual bounds: Verifying statements~\ref{statement-4} and~\ref{statement-5} of Section~\ref{sec:overview-TD}}

Rebalancing a cut~$\scrT$ with~$\ell$ trees creates no more than~$\ell+2$ new~$3$-full nodes (the~$\ell-1$ ancestors of~$\calT$ and the parent, grand-parent and great-grand-parent of the resulting tree), and no more than~$\ell$ new~$\{1,2\}$-nodes (the~$\ell-1$ ancestors of~$\calT$ and the parent of the resulting tree).

Then, each batch of operations performed during an insertion consists in:

\begin{itemize}[leftmargin=4.8mm,itemsep=0.5pt,topsep=0.5pt]
\item calling~$\sfF\sfE$, which rebalances a cut with~$8$ trees, thus creating no more than~$10$ new~$3$-full nodes and~$8$ new~$\{1,2\}$-nodes, and makes~$x^{d}$ a~$\{0,1\}$-node;

\item promoting the~$\{0,1\}$-nodes~$x^i$ (where~$2 \leqslant i \leqslant d$); this transforms~$x^i$ it into a~$\{1,2\}$-node, and turns the~$\{1,1\}$-node~$x^{i-1}$ into a~$\{0,1\}$-node, i.e., it creates one~$\{1,2\}$-node and destroys a~$3$-full node;

\item promoting~$x^1$ and, if necessary, performing a rotation or rebalancing operation; this amounts to rebalancing a cut with at most~$16$ trees, thus creating no more than~$18$ new~$3$-full nodes and~$16$ new~$\{1,2\}$-nodes.
\end{itemize}

Consequently, this batch increases the number of~$3$-full nodes by no more than~$27 - d$, and the number of~$\{1,2\}$-nodes by no more than~$25 + d$: this is statement~\ref{statement-4}.

Similarly, each batch of operations performed during a deletion consists in:

\begin{itemize}[leftmargin=4.8mm,itemsep=0.5pt,topsep=0.5pt]
\item calling~$\sfN\sfS$ if necessary, which rebalances a cut with no more than~$11$ nodes, thus creating no more than~$13$ new~$3$-full nodes and~$11$ new~$\{1,2\}$-nodes, and makes~$x^{d}$ a four-node;

\item save the nodes~$x^i$ (where~$2 \leqslant i \leqslant d$) from being four-nodes, at the expense of making~$x^{i-1}$ a four-node instead; this requires a rotation or demotion prescribed by cases~\ref{sota:case-6} to~\ref{sota:case-8} of the textbook algorithm, creates no~$3$-full node, and destroys at least one~$\{1,2\}$-node;

\item save~$x^1$ from being a four-node, at the expense of making~$x^0$ a~$\{1,2\}$-node instead of a~$\{1,1\}$-node, or or performing the simple rotation of case~\ref{sota:case-5} of the textbook algorithm, in case~$x^0$ was a~$\{2,2,2\}$-node;
this creates no~$3$-full nodes and no more than~$2$ new~$\{1,2\}$-nodes.
\end{itemize}

Consequently, this batch increases the number of~$3$-full nodes by no more than~$11$, and the number of~$\{1,2\}$-nodes by no more than~$14 - d$: this is statement~\ref{statement-5}.

\bibliography{AVL}

\end{document}